\documentclass[english,aps,prd,showpacs,superscriptaddress,nofootinbib]{revtex4}
\usepackage{amsfonts}
\usepackage{amsmath}
\usepackage{amssymb}
\usepackage{appendix}
\usepackage{babel}
\usepackage{bm}
\usepackage{bbm}
\usepackage{braket}
\usepackage{color}
\usepackage{float}
\usepackage{fouridx}
\usepackage[T1]{fontenc}
\usepackage{epsfig}
\usepackage{graphicx}
\usepackage[colorlinks,linkcolor=blue]{hyperref}
\usepackage{mathrsfs}
\usepackage{multirow}
\usepackage{slashed}
\usepackage{subfigure}
\usepackage{verbatim}
\usepackage{xcolor}
\usepackage{ulem}
\newcommand {\bseq}{\begin{subequations}}
\newcommand {\eseq}{\end{subequations}}
\newcommand{\bqa}{\begin{eqnarray}}
\newcommand{\eqa}{\end{eqnarray}}
\newcommand{\beq}{\begin{equation}}
\newcommand{\eeq}{\end{equation}}
\newcommand{\nn}{\nonumber}

\allowdisplaybreaks[0]
\begin{document}

\title{Inclusive ${\bm h}_{\bm c}$ production and energy spectrum from
${\bm e}^{\bm +}{\bm e}^{\bm -}$ annihilation at Super $\bm{B}$ factory}

\author{Qing-Feng Sun\footnote{qfsun@mail.ustc.edu.cn}}
	\affiliation{Department of Modern Physics, University of Science and Technology of China, Hefei, Anhui 230026,  China\vspace{0.2cm}}
\affiliation{Institute of High Energy Physics, Chinese Academy of Sciences, Beijing 100049, China\vspace{0.2cm}}

\author{Yu Jia\footnote{jiay@ihep.ac.cn}}
	\affiliation{Institute of High Energy Physics, Chinese Academy of Sciences, Beijing 100049, China\vspace{0.2cm}}
	
\author{Xiaohui Liu\footnote{xiliu@bnu.edu.cn}}
	\affiliation{Center of Advanced Quantum Studies and Department of Physics, Beijing Normal University, Beijing, 100875, China\vspace{0.2cm}}

\author{Ruilin Zhu\footnote{rlzhu@njnu.edu.cn}}
	\affiliation{Department of Physics and Institute of Theoretical Physics, Nanjing Normal University, Nanjing, Jiangsu 210023, China\vspace{0.2cm}}
	
\date{\today}

\begin{abstract}
We calculate the next-to-leading order (NLO) radiative correction to the color-octet $h_c$ inclusive
production in $e^+e^-$ annihilation at Super $B$ factory, within the nonrelativistic QCD factorization framework.
The analytic expression for the NLO short-distance coefficient (SDC) accompanying the color-octet production operator $\mathcal{O}_8^{h_c}(^1S_0)$ is obtained after summing both virtual and real corrections.
The size of NLO correction for the color-octet production channel is found to be positive and substantial.
The NLO prediction to the $h_c$ energy spectrum is plagued with unphysical endpoint singularity.
With the aid of the soft-collinear effective theory, those large endpoint logarithms are resummed to the
next-to-leading logarithmic (NLL) accuracy. Consequently, further
supplemented with the non-perturbative shape function,
we obtain the well-behaved predictions for the $h_c$ energy spectrum in the entire kinematic range,
which awaits the examination by the forthcoming \textsf{Belle II} experiment.


%\keywords{Perturbative calculations, Heavy quarkonia, Nonrelativistic QCD, Soft-collinear effective theory}

\end{abstract}

\maketitle

\section{Introduction}

The $h_c(1P)$ meson, the lowest-lying spin-singlet $P$-wave charmonium, is the last member found among the
charmonium family below the open charm threshold. The first hint about its existence was reported in the
process $p\bar{p}\to h_c \to J/\psi\pi^0$ by the Fermilab \textsf{E760} experiment in 1992~\cite{Armstrong:1992ae}.
Finally, in 2005, the $h_c$ state was firmly established through the process $p\bar{p}\to h_c\to \eta_c\gamma$ in the
Fermilab \textsf{E835} experiment~\cite{Andreotti:2005vu}, as well as through the
isospin-violating charmonium transition process $\psi(2S)\to h_c (\to \eta_c\gamma) +\pi^0$ in the
\textsf{CLEO-c} experiment~\cite{Rosner:2005ry,Rubin:2005px}.
Later this decay chain was confirmed in the \textsf{BESIII} experiment
with much greater data sample~\cite{Ablikim:2010rc,Ablikim:2012ur}.
To date, the latest measured mass and width of $h_c$ are $M_{h_c}= 3525.38\pm 0.11$ MeV, and
$\Gamma_{h_c}= 0.7\pm 0.28\pm 0.22$ MeV, respectively~\cite{Olive:2016xmw}.
Two exclusive decay channels, the electric dipole ($E1$) radiative
transition $h_c\to \eta_c\gamma$,  and the OZI-suppressed annihilation decay $h_c\to 2\pi^+ 2\pi^- \pi^0$, have been measured,
with the corresponding branching fractions ${\mathcal B}(h_c\to \eta_c\gamma)=(51\pm6)\%$,
and ${\mathcal B}(h_c\to 2\pi^+ 2\pi^- \pi^0)=(2.2^{+0.8}_{-0.7})\%$, respectively~\cite{Olive:2016xmw}.
It is worth mentioning that, the ${}^1P_1$ counterparts in the bottomonium family,
the $h_b(1P, 2P)$ mesons, have also recently been established via the di-pion transition from the
$\Upsilon(5S)$ resonance in the \textsf{Belle} experiment~\cite{Adachi:2011ji}.

It is interesting to ask whether one can possibly understand various dynamical aspects of the $h_c$ meson from the first principles of QCD.
In fact, nonrelativistic QCD (NRQCD)~\cite{Caswell:1985ui}, the modern effective field theory to describe the slowly-moving heavy quark-antiquark system, is an appropriate model-independent framework to tackle a multi-scale system exemplified by the charmonium state $h_c$.
Furthermore, the NRQCD factorization approach~\cite{Bodwin:1994jh}, originally developed by Bodwin, Braaten and Lepage, provides a powerful and systematic language to describe the inclusive quarkonium production and decay processes,
which has been fruitfully applied to uncountable charmonium phenomenologies in the past two decades~\cite{Brambilla:2010cs}.

For the dominant $E1$ decay process $h_c\to \eta_c\gamma$, there have been many preceding
studies based on the multipole expansion picture in the quark potential models~\cite{Novikov:1977dq}.
Moreover, the radiative and relativistic corrections to the inclusive hadronic widths of $h_{c,b}$ have recently
been investigated in the NRQCD factorization framework~\cite{Li:2012rn}.
On the other hand, the $h_c$ production in various collision environments have also been extensively investigated in
recent years. For instance, $h_c$ inclusive production in $B$ meson decay~\cite{Bodwin:1992qr, Beneke:1998ks}, $h_c$ photoproduction~\cite{Fleming:1998md},
$h_c$ hadroproduction~\cite{Sridhar:2008sc,Qiao:2009zg,Wang:2014vsa}, inclusive $h_c$ production from $e^+e^-$
annihilation~\cite{Jia:2012qx, Wang:2012tz},
exclusive $h_c$ production from $Z^0$ decay~\cite{Chen:2013itc}, from double charmonium production in $e^+e^-$ annihilation~\cite{Chen:2015zta},
as well as from $\Upsilon(nS)$ decay~\cite{Zhu:2015jha}.

The hadroproduction rate of $h_c$ is significant at LHC experiment due to the huge partonic luminosity.
A recent computation indicates that the gluon-to-$h_c$ fragmentation probability may reach the order $10^{-6}$~\cite{Feng:2017cjk}.
In sharp contrast to $J/\psi(\psi^\prime)$ hadroproduction~\cite{Abulencia:2007us,Aaij:2011jh,Aad:2011sp,Khachatryan:2010yr,Abelev:2011md},
unfortunately it is rather challenging to reconstruct the $h_c$ events via the dominant decay channel
$h_c\to \eta_c\gamma$, due to the tremendous background at hadron collision experiments.
In contrast, tagging $h_c$ is much more tractable in the $e^+e^-$ machines than in hadron colliders.
For example, the exclusive $h_c$ production process $e^+e^-\to h_c \pi^+ \pi^-$ at center-of-mass energy $\sqrt{s}=4.170$~GeV
has been studied by the \textsf{CLEO} Collaboration, with the cross section measured to be $15.6\pm2.3\pm1.9\pm3.0$~pb~\cite{CLEO:2011aa}.
They also found evidence for the process $e^+ + e^-\to h_c \eta$ at $3\sigma$ confidence level.
As a byproduct of studying this exclusive $h_c$ production channel, \textsf{BESIII} have recently found two
charmonium-like resonances, namely the $Y(4220)$ and $Y(4390)$~\cite{BESIII:2016adj}.

The forthcoming \textsf{Belle II} experiment (also referred to as Super $B$ factory) will accumulate a
tremendous dataset near the $\Upsilon(4S)$ energy.
In this paper, we will focus on the inclusive $h_c$ production in $e^+e^-$ annihilation at $\sqrt{s}\approx 10.58$ GeV, near the
$\Upsilon(4S)$ resonance.
In the previous work~\cite{Jia:2012qx, Wang:2012tz}, the NRQCD SDCs were evaluated for both color-singlet and color-octet channels
at the leading order (LO) in $\alpha_s$, and it was found that the latter octet-channel production cross section dominated the singlet-channel cross section.
Therefore, in order to make a more precise prediction, it is helpful to evaluate the NLO QCD correction to the color-octet cross section.
Moreover, to expedite the experimental search for $h_c$, it is crucial to predict not only the total $h_c$ production rate,
but also the differential $h_c$ energy spectrum.

The LO color-octet contribution to the $h_c$ energy spectrum in $ e^+ e^- \to h_c+X $ is simply a $ \delta$-function,
determined by the partonic process $e^+e^- \to c\bar{c}({}^1S_0^{(8)})+g$.
After including the real correction in the color-octet channel, $e^+e^- \to c\bar{c}({}^1S_0^{(8)})+gg$,
the energy spectrum then becomes continuous over all allowed domain, however turns out to be singular near the upper endpoint,
due to the soft and collinear gluon radiation in this limited region of phase space.
This signals a breakdown of the fixed-order QCD prediction, and failure of NRQCD expansion near this kinematic endpoint region.
The aim of this work is thus two fold. First we extend the LO color-octet NRQCD SDC obtained in \cite{Jia:2012qx} to
NLO in $\alpha_s$, in a fully analytical manner.
Secondly, we follow the recipe of the resumming large logarithms in the color-octet channel for the process $e^+e^-\to J/\psi+X$ near the endpoint region~\cite{Fleming:2003gt}, which was formulated in the context of the soft-collinear effective theory (SCET)~\cite{Bauer:2000ew,Bauer:2000yr,Bauer:2001ct,Bauer:2001yt,Bauer:2002nz,Beneke:2002ph}, to tame the endpoint singularity encountered in our case, and finally predict a well-behaved $h_c$ energy spectrum.
We hope our prediction will provide some useful guidance for unambiguously erecting the $h_c$ state
in the forthcoming \textsf{Belle II} experiment.

The rest of the paper is distributed as follows.
%-----------------------------
In Sec.~\ref{sec:fixed-order calculation}, the fixed-order calculations for the SDCs are presented within the NRQCD factorization framework.
We first review the existing LO results for both color-singlet and octet channels.
%-----------------------------
In Sec.~\ref{Order:alphas:corr},
we present the analytical expressions for NLO NRQCD SDCs from the color-octet channel, including both virtual and real corrections.
%-----------------------------
In Sec.~\ref{sec:resummation}, within the SCET framework, we show how to resum the large endpoint logarithms to the NLL accuracy.
%-----------------------------
In Sec.~\ref{sec:numerical results}, we present our numerical results for the total $h_c$ production rate and its differential energy spectrum.
We also discuss the observational prospects of the $h_c(1P,2P)$ states in the forthcoming \textsf{Belle II} experiment.
%-----------------------------
Finally we summarize in Sec.~\ref{sec:summary}.
%-----------------------------
In the Appendix, we expound how to analytically carry out the three-body phase space to
isolate the soft and collinear divergences in $d=4-2\epsilon$ spacetime dimensions.
%-----------------------------

%%%%%%%%%%%%%%%%%%%%%%%%%%%%%%%%%%%%%%%%%%%%%%%%%%
\section{NRQCD factorization and LO Short-distance coefficients
\label{sec:fixed-order calculation}}
%%%%%%%%%%%%%%%%%%%%%%%%%%%%%%%%%%%%%%%%%%%%%%%%%%

\subsection{NRQCD factorization for $h_c$ production}

Heavy quarkonium is a QCD bound state predominantly composed of a pair of nonrelativistic heavy quark and antiquark.
For the charmonium, the typical velocity between the charm quarks inside a charmonium is roughly $v^2\approx 0.3$,
thus velocity expansion is not anticipated to converge very well.
According to the NRQCD factorization theorem~\cite{Bodwin:1994jh}, the inclusive production rate of $h_c$ can be expressed
as a sum of the product of perturbatively calculable NRQCD SDCs and the non-perturbative NRQCD long-distance matrix elements (LDMEs).
The importance of the LDMEs is weighed by the power counting in $v$.
At the lowest order in $v$, the differential cross section for inclusive $h_c$ production can be written as~\cite{Bodwin:1994jh}
%-----------------------
\beq
%-----------------------
\label{NRQCD formalism}
%-----------------------
d \sigma[e^{+}e^{-}\to h_c+X]=\frac{dF_1(\mu_\Lambda)}{m_c^4} \langle\mathcal{O}_{1}^{h_c}(^1P_1)\rangle+ \frac{dF_8}{m_c^2}
\langle\mathcal{O}_{8}^{h_c}(^1S_0)(\mu_\Lambda)\rangle+\cdots,
%-----------------------
\eeq
%-----------------------
where the SDCs $dF_1$ and $d F_8$ can be calculated order by order in $\alpha_s$,
$\langle\mathcal{O}_{1}^{h_c}(^1P_1)\rangle$ and $\langle\mathcal{O}_{8}^{h_c}(^1S_0)\rangle$ are the
color-singlet and color-octet NRQCD production LDMEs, respectively.
The corresponding $h_c$ production operators in NRQCD are defined as~\cite{Bodwin:1994jh}~\footnote{
It was first made clear by Nayak, Qiu and Sterman~\cite{Nayak:2005rw,Nayak:2005rt} a decade ago that the
original definition of the NRQCD color-octet production operator~\cite{Bodwin:1994jh}
is not gauge invariant, and the correct definition necessitates the inclusion of
eikonal lines that run from the location of the quark/antiquark fields to infinity.
To the perturbative order considered in this work,
this nuisance does not play a role so we adhere to the conventional definition~\cite{Bodwin:1994jh}.}
%----------------------
\begin{subequations}
%----------------------
\label{eq:NRQCD production operators}
%----------------------
\begin{align}
%----------------------
\mathcal{O}_{1}^{h_c}(^1P_1)  =&
\chi^\dagger \left(-\frac{i}{2}\tensor{\bf{D}}\right) \psi  \sum_{X} |h_c+X\rangle \cdot\langle h_c+X|
\psi^\dagger \left(-\frac{i}{2}\tensor{\bf{D}}\right) \chi,
%----------------------
\\
%----------------------
\mathcal{O}_{8}^{h_c}(^1S_0) =&
\chi^\dagger T^a\psi \sum_{X} |h_c+X\rangle\cdot \langle h_c + X|\; \psi^\dagger T^a\chi,
%----------------------
\end{align}
%----------------------
\end{subequations}
%----------------------
where $\psi$ and $\chi$ denote the Pauli spinor fields that annihilates a heavy quark and creates a heavy antiquark, respectively.
$\tensor{\bf{D}}$ represents the left-right symmetric spatial component of the covariant derivative
$D_\mu=\partial_\mu - i g_s T^a A_\mu^a$, and $T^a$ ($a=1,\ldots,8$) signifies the generator
in the fundamental representation of the $SU(3)_c$ group.
The $\mu_\Lambda$ refers to the NRQCD factorization scale, which lies in the range $m_c v \leq \mu_\Lambda \leq m_c$.
These two NRQCD production operators are interconnected through the NRQCD
renormalization group equation (RGE)~\cite{Bodwin:1994jh}:
%----------------------
\beq
%----------------------
\frac{d}{d\ln\mu^2_\Lambda}\langle {\cal O}_8^{h_c}(^1S_0)(\mu_\Lambda)\rangle =
\frac{2 C_F \alpha_s(\mu_{\Lambda}) }{3 \pi N_c m_c^2} \langle {\cal O}_1^{h_c}(^1P_1)\rangle.
%----------------------
\eeq
%----------------------
Being infrared-finite, $dF_1$ and $dF_8$ are insensitive to the long-distance hadronization effects,
thus can be determined through the standard perturbative matching procedure. One can replace the physical $h_c$ state in Eq.~\eqref{NRQCD formalism} by the free on-shell $c\bar{c}$ pairs with quantum numbers ${}^1S_0^{(8)}$ or ${}^1P_1^{(1)}$, computing both sides of Eq.~\eqref{NRQCD formalism}, demanding both perturbative QCD and perturbative
NRQCD to generate identical results. Ultimately, one can solve these two linear equations to ascertain the two
SDCs, order by order in $\alpha_s$.
Here we stress that it is crucial to include the color-octet contribution, otherwise the uncancelled IR divergences emerging from the color-singlet channel would impede the predictive power of NRQCD.
For the computation in the QCD side, it is convenient to employ the covariant
projection technique~\cite{Petrelli:1997ge,Bodwin:2010fi}
to project the $c\bar{c}$ amplitude onto the intended ${}^{2S+1}L_J$ states.
Throughout this work, Dimensional Regularization (DR), that is, to work in the spacetime dimensions
$d=4-2\epsilon$, is adopted to regularize both UV and IR divergences.

A kinematical simplification also arises from the $s$-channel nature of this process.
As long as we are concerned only with the $h_c$ energy distribution,
one can reexpress the $h_c$ production rate from $e^+e^-$ annihilation
in terms of that from virtual photon decay~\cite{Keung:1980ev}:
%----------------------
\beq
%----------------------
\label{cross section}
%----------------------
d\sigma\left[e^{+}e^{-} \to h_c+X\right]=\frac{4\pi \alpha}{s^{3/2}} d\Gamma\left[\gamma^{*} \to h_c+X\right],
%----------------------
\eeq
%----------------------
where the center-of-mass energy of the $e^{+}e^{-}$ system is denoted by $\sqrt{s}$.

\begin{figure}[tb]
	\centering
	\includegraphics[width=1.0\textwidth]{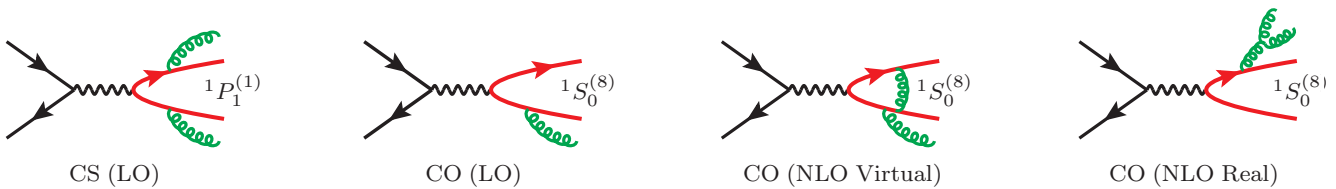}
	\caption{Representative Feynman diagrams for the $c\bar{c}(n)$ production
from $e^+e^-$ annihilation, for $n={}^1S_0^{(8)}$
or ${}^1P_1^{(1)}$.}
\label{fig:feynman diagrams}
\end{figure}

Some representative Feynman diagrams for $c\bar{c}(n)$ ($n={}^1S_0^{(8)}$
or ${}^1P_1^{(1)}$) production from $e^+e^-$ annihilation in both
color-singlet and color-octet channels are shown in Fig.~\ref{fig:feynman diagrams}.
Due to the odd $C$ parity of the $h_c$ meson, the color-singlet channel starts at $\mathcal{O}(\alpha^2_s)$, while the octet
contribution starts at $\mathcal{O}(\alpha_s)$.
In the rest of this section, we will briefly review the LO results for $F_8$ and $F_1$, which were first analytically evaluated in Ref.~\cite{Jia:2012qx}.

\subsection{LO color-octet SDC}
\label{sec:LO color octet}

At LO in color-octet channel, we only need consider $e^+e^-\to \gamma^* \to  c\bar{c}({}^1S_0^{(8)})+g$.
The differential two-body phase space in $d=4-2\epsilon$ dimensions reads~\cite{Jia:2012qx}
%-----------------------
\beq
%-----------------------
\label{eq:Phi2}
%-----------------------
d\Phi_2=\frac{c_{\epsilon}}{8\pi}s^{-\epsilon}(1-r)^{1-2\epsilon}\delta(1+r-z) \,dz,
%-----------------------
\eeq
%-----------------------
where
%-----------------------
\beq
%-----------------------
\label{Def:cepsilon:and:r}
%-----------------------
c_{\epsilon}\equiv(4\pi)^{\epsilon}\frac{\Gamma(1-\epsilon)}{\Gamma(2-2\epsilon)},\qquad
r\equiv {4m_c^2 \over s},\qquad z\equiv {2P^0 \over \sqrt{s}},
%-----------------------
\eeq
%-----------------------
with $P^{\mu}=(P^0,\bf{P}) $ representing the four-momentum of the $c\bar{c}$ pair.
%\sout{In light of energy conservation, the energy fraction $z$ is restricted to
%	the range $2\sqrt{r}\le z \le 1+r$.}

The LO amplitude squared turns to be
%-----------------------
\begin{align}
%-----------------------
\label{eq:LO amplitude squared}
%-----------------------
&\sum_{\text{Pol,Col}}{\left|\mathcal{M}^{(0)}\left[\gamma^{*}\to c\bar{c}(\fourIdx{1}{}{(8)}{0}{S})+g\right]\right|^2}
%-----------------------
% \nonumber\\
%-----------------------
=256\pi^2 e_c^2 \alpha C_A C_F \alpha_s \mu_r^{2\epsilon} (1-\epsilon)(1-2\epsilon),
%-----------------------
\end{align}
%-----------------------
where $e_c=\frac{2}{3} $ is the electric charge of the charm quark,
$C_A=3$ and $C_F={4\over 3}$ are the Casimirs of the color $SU(3)_c$ group.
Integrating Eq.~\eqref{eq:LO amplitude squared} over the two-body phase space in Eq.~\eqref{eq:Phi2}, we obtain
%-----------------------
\begin{align}
%-----------------------
\hat{\sigma}_\text{LO}^{(8)}\equiv &\frac{2\pi \alpha}{3s^2}\int{d\Phi_2}\,\sum_{\text{Pol,Col}}{\left|\mathcal{M}^{(0)}\left[\gamma^{*}\to c\bar{c}(\fourIdx{1}{}{(8)}{0}{S})+g\right]\right|^2}
%-----------------------
\nn\\
%-----------------------
=& \left(\frac{4\pi\mu_r^2}{s}\right)^{\epsilon}\frac{\Gamma(2-\epsilon)}{\Gamma(1-2\epsilon)} \frac{64\pi^2 e_c^2\alpha^2 C_A C_F \alpha_s(1-r)^{1-2\epsilon}}{3s^2}.
%-----------------------
\label{hat:sigma:8:LO}
\end{align}
%-----------------------
The factor $\frac{1}{3} $ accounts for averaging over the three polarizations of $\gamma^{*}$. The differential expression of $ \hat{\sigma}_\text{LO}^{(8)} $ in $ 4 $ dimensions reads
%-----------------------
\beq
%-----------------------
\left.\frac{d\hat{\sigma}_\text{LO}^{(8)}}{d z}\right|_{d\to 4}=
\frac{64\pi^2 e_c^2\alpha^2 C_A C_F \alpha_s(1-r)}{3s^2}\delta(1+r-z).
%-----------------------
\label{LO:CO:diff:parton:cross:sec}
\eeq
%-----------------------

Substituting Eq.~\eqref{LO:CO:diff:parton:cross:sec} to the left hand side of Eq.~\eqref{NRQCD formalism},
and only retaining the NRQCD matrix element in the color-octet channel,
we then deduce the LO color-octet SDC:
%-----------------------
\beq
%-----------------------
\label{eq:dF8 LO}
%-----------------------
{d F_8^{\rm LO} \over dz} =
{m_c \over \langle\mathcal{O}_8^{c\bar{c}}({}^1S_0)\rangle}
{d \hat{\sigma}_\text{LO}^{(8)}\over d z}
%-----------------------
 = {64\pi^2e_c^2\alpha^2 C_A C_F \alpha_s(1-r)m_c \over 3(N_c^2-1) s^2}\delta(1+r-z),
%-----------------------
\eeq
%-----------------------
where we have used $\langle\mathcal{O}_8^{c\bar{c}}({}^1S_0)\rangle=N_c^2-1$.
The integrated color-octet SDC is then
%-----------------------
\beq
%-----------------------
\label{eq:F8 LO}
%-----------------------
F_8^{\rm LO}= {64\pi^2e_c^2\alpha^2 C_A C_F\alpha_s(1-r)m_c \over 3(N_c^2-1) s^2},
\eeq
%-----------------------
which scales as $1/s^2$ asymptotically.

\subsection{LO color-singlet SDC}
\label{sec:LO color singlet}

To determine the LO SDC in the color-singlet channel, we need consider the partonic process $e^+e^-\to c\bar{c}({}^1P_1^{(1)})+gg$.
The IR divergence appears in the upper endpoint of the $h_c$ spectrum, when one of the gluons becomes soft.
It is most convenient to handle this IR singularity using DR.
As a virtue of the color-octet mechanism of NRQCD, the single IR pole is factored into the color-octet NRQCD LDME.
As a remnant of this IR divergence, the
renormalized color-octet LDME is defined at the NRQCD factorization scale $\mu_\Lambda$,
in the meanwhile the SDC $F_1$ acquires an explicit
logarithmic dependence on $\mu_\Lambda$.
The differential color-singlet SDC $d F_1/d z$ is somewhat too lengthy to reproduce here, and
we refer the interested readers to Ref.~\cite{Jia:2012qx} for its complete expression.
Here we just present the integrated color-singlet SDC:
%-----------------------
\begin{align}
%-----------------------
\label{eq:F1 LO}
%-----------------------
F_1^{\text{LO}}(\mu_\Lambda) = &\frac{64\pi e_c^2\alpha^2 C_F \alpha_s^2 m_c}{9N_c s^2}(1-r)\left[-\ln{\frac{\mu_{\Lambda}^2}{4m_c^2}}+2\ln{(1-r)}-\frac{65-84r}{12(1-r)}
+\frac{7+7r-9r^2}{6(1-r)^2}\ln{r}\right.
%-----------------------
\nn\\
%-----------------------
&\left.+\frac{r(5-7r)}{16(1-r)^2}\ln^2{\frac{1+\sqrt{1-r}}{1-\sqrt{1-r}}}+
\frac{14-15r}{8(1-r)^{3/2}}\ln{\frac{1+\sqrt{1-r}}{1-\sqrt{1-r}}}\right],
%-----------------------
\end{align}
%-----------------------
which is obtained according to the $\overline{\rm MS}$ renormalization scheme.
It is enlightening to see the asymptotic behavior of $F_1^{\text{LO}}$ in
the $ \sqrt{s}\gg m_c $ limit:
%-----------------------
\beq
%-----------------------
\label{eq:Asymptotic behavior of F1 at LO}
%-----------------------
F_1^{\text{LO}}(\mu_\Lambda)\Big|_{\rm Asym.} =
\frac{64\pi e_c^2\alpha^2 C_F \alpha_s^2 m_c}{9N_c
s^2}\left(-\frac{7}{12}\ln{r}-\ln{\frac{\mu_\Lambda^2}{4m_c^2}}
%-----------------------
%-----------------------
-\frac{65}{12}+\frac{7}{2}\ln{2}\right),
%-----------------------
\eeq
%-----------------------
which is proportional to $1/s^2$ times a single logarithm of $s/m_c^2$.

\section{NLO radiative correction for the color-octet channel}
\label{Order:alphas:corr}

In this section, we are going to calculate the NLO radiative correction for the color-octet SDC $d F_8$,
which includes the real correction $e^+e^-\to c\bar{c}({}^1S_0^{(8)})+gg(q\bar{q})$,
together with the one-loop virtual correction to $e^+e^-\to c\bar{c}({}^1S_0^{(8)})+g$.
The UV divergences encountered in virtual correction will be eliminated
by the standard renormalization procedure,
while the IR singularities turn out to cancel out
after summing both real and virtual corrections.

In the NLO calculation, we generate the QCD Feynman diagrams and amplitudes using the package
\textsf{FeynArts}~\cite{Hahn:2000kx},
and employ the package \textsf{FeynCalc}~\cite{Shtabovenko:2016sxi} to carry out contraction of the Lorentz
indices and trace over Dirac matrices.
We use the Feynman gauge throughout the calculation.

\subsection{Real correction}

There are more Feyman diagrams for $e^+e^-\to \gamma^*\to c\bar{c}({}^1S_0^{(8)})+gg$ than the color-singlet channel,
since the three-gluon vertex is permitted due to the color-octet feature of
the $c\bar{c}$ pair.
Furthermore, the new channel $e^+e^-\to \gamma^*\to c\bar{c}({}^1S_0^{(8)})+q\bar{q}$ also becomes permissible.
One typical real emission diagram is depicted in Fig.~\ref{fig:feynman diagrams}.

In this section, we will quickly present the analytic results by integrating the squared amplitudes
over the three-body phase space in DR, closely following the recipe outlined in Ref.~\cite{Jia:2012qx}.
To ensure the correctness of our results, we also redo the calculation using numerical recipe,
{\it i.e.}, utilizing the two-cutoff phase space slicing method~\cite{Harris:2001sx},
and find full agreement with our analytical results.

First, let us introduce, in addition to $z$, two additional fractional energy variables, $x_1 $ and $ x_2 $:
%-----------------------
\beq
%-----------------------
x_1\equiv \frac{2k_1^0}{\sqrt{s}}, \qquad x_2\equiv \frac{2k_2^0}{\sqrt{s}},
%-----------------------
\eeq
%-----------------------
where $ k_1 $ and $ k_2 $ represent the momenta of the final-state gluons (or light quark and antiquark)
in real emission process.
These variables are subject to the constraint $x_1+x_2+z=2 $ by energy conservation.

For convenience, we separate the squared amplitude for $ \gamma^{*}\to c\bar{c}(\fourIdx{1}{}{(8)}{0}{S})+gg $
into four pieces:
%-----------------------
\beq
%-----------------------
\label{eq:amplitude squared for 1S08 gg}
%-----------------------
\sum_{\text{Pol,Col}}{\left|{\mathcal M}^{\rm R}\left[\gamma^{*}\to c\bar{c}(\fourIdx{1}{}{(8)}{0}{S})+gg\right]\right|^2}=\mathcal{I}_{\text{S}}(x_i,z)
+\mathcal{I}_{\text{C}}(x_i,z)+\mathcal{I}_{\text{SC}}(x_i,z)+\mathcal{I}_{\text{Fin}}(x_i,z),
%-----------------------
\eeq
%-----------------------
with $i=1,2$. Explicitly, these four pieces are
%-----------------------
\bseq
%-----------------------
\label{eq:four parts}
%-----------------------
\begin{align}%\label{soft and collinear terms}
%-----------------------
\mathcal{I}_{\text{S}}(x_i,z)=& -\frac{2^{12}\pi^3 e_c^2\alpha C_A^2 C_F \alpha_s^2 \mu_r^{4\epsilon}\left(1-\epsilon\right)\left(1-2\epsilon\right)r}{s}\left[\frac{1}{\left(1+r-z-x_1\right)^2}+\frac{1}{\left(1+r-z-x_2\right)^2}\right],
%-----------------------
\label{eq:soft term}
\\
%-----------------------
{\cal I}_{\rm C}(x_i,z)=&{2^{12}\pi^3 e_c^2\alpha C_A^2 C_F \alpha_s^2\mu_r^{4\epsilon}(1-\epsilon)
(1-2\epsilon)\over (1-r)^2 s} {x_1 x_2 - 2(1-r)^2 \over 1+r-z},
%%-----------------------
\label{eq:collinear term}
\\
%-----------------------
\mathcal{I}_{\text{SC}}(x_i,z)=& -\frac{2^{12}\pi^3 e_c^2\alpha C_A^2 C_F \alpha_s^2\mu_r^{4\epsilon}\left(1-\epsilon\right)\left(1-2\epsilon\right)(1-r)}{s}
\frac{1}{1+r-z}\left(\frac{1}{1+r-z-x_1}+\frac{1}{1+r-z-x_2}\right),
%-----------------------
\label{eq:soft and collinear term}
\\
%%-----------------------
 {\cal I}_{\rm Fin}(x_i,z)=&
 {2^{12} \pi^3 e_c^2 \alpha C_A^2 C_F\alpha_s^2 \over (1-r)^2 (2-z)^2 (1-r-x_1)^2 (1-r-x_2)^2 s}
 \bigg\{(1-r)^3(1+r-z)(z-2r)(2-z)^2
%-----------------------
\nn\\
%-----------------------
&- (1-r)^3 (5+2r+r^2-5z-r z+ z^2) x_1 x_2 + 2 (1-r) (2-z)^2 x_1^2 x_2^2 - (3-r-z) x_1^3 x_2^3\bigg\}.
%-----------------------	
%-----------------------
\label{eq:finite term}
%-----------------------
\end{align}
%-----------------------
\eseq
%-----------------------
Each individual term is symmetric under the exchange $x_1\leftrightarrow x_2$, reflecting the Bose symmetry of
the two gluons in the final state.
Upon phase space integration, the first term $ \mathcal{I}_{\text{S}}$
would lead to a single soft pole, when one of the gluons becomes soft.
The second term $ \mathcal{I}_{\text{C}}$ would result in a
single collinear pole, when the final-state gluons become collinear to each other.
The third term $\mathcal{I}_{\text{SC}}$ would produce the double IR pole,
arising from the corner of phase space where one of the gluons becomes simultaneously soft and collinear to the other one.
Note both soft and collinear singularities can arise only when the $c\bar{c}$ pair acquires its maximal energy,
that is, in the $z\to 1+r$ limit.
The last term $\mathcal{I}_{\text{Fin}}$ will not result in any IR divergences upon phase space
integration, therefore can be directly treated in 4 spacetime dimensions.

Integrating the squared amplitudes in Eq.~\eqref{eq:amplitude squared for 1S08 gg} over the three-body phase space,
we obtain
%-----------------------
\beq
%-----------------------
\hat{\sigma}_{\text{R}}^{(8),gg}\equiv \hat{\sigma}_{\text{Div}}^{(8),gg}+\hat{\sigma}_{\text{Fin}}^{(8),gg},
%-----------------------
\label{eq:decomposing:X:section}
%-----------------------
\eeq
%-----------------------
where the ``divergent'' and ``finite'' partonic cross sections are defined as
%-----------------------
\bseq
%-----------------------
\label{eq:gg sigma hat}
%-----------------------
\begin{align}
%-----------------------
\hat{\sigma}_{\text{Div}}^{(8),gg}=&\int_{2\sqrt{r}}^{1+r}d z\frac{d\hat{\sigma}_{\text{Div}}^{(8),gg}}
{dz}=\frac{1}{2!}\frac{2\pi \alpha}{3s^2}\int{d\Phi_3}\,\left[\mathcal{I}_{\text{S}}\left(x_i,z\right)+
\mathcal{I}_{\text{C}}(x_i,z)+\mathcal{I}_{\text{SC}}(x_i,z)\right],
%-----------------------
\\
%-----------------------
\hat{\sigma}_{\text{Fin}}^{(8),gg}=&\int_{2\sqrt{r}}^{1+r}d z\frac{d\hat{\sigma}_{\text{Fin}}^{(8),gg}}{dz}=\frac{1}{2!}\frac{2\pi \alpha}{3s^2}\int{d\Phi_3}\,\mathcal{I}_{\text{Fin}}\left(x_i,z\right).
%-----------------------
\end{align}
%-----------------------
\eseq
%-----------------------
Here $d\Phi_3$ signifies the three-body phase space measure, whose exact definition in $d=4-2\epsilon$
dimensions is given in Eq.~\eqref{eq:3 body phase space}.
We have also included a symmetry factor ${1\over 2!}$  in Eq.~\eqref{eq:gg sigma hat},
to account for the indistinguishability of the final-state gluons.

By carrying out one-fold integration over $x_1$ in Eq.~\eqref{eq:gg sigma hat},
we then arrive at the partonic cross section differential in the energy fraction of the $c\bar{c}$ pair:
%-----------------------
\bseq
%-----------------------
\label{eq:differential cross section for gg8}
%-----------------------
\begin{align}
%-----------------------
\frac{d\hat{\sigma}_{\text{Div}}^{(8),gg}}{d z} =&\hat{\sigma}_\text{LO}^{(8)}\frac{\alpha_s}{\pi}\frac{(1-r)^{-2\epsilon}r^{\epsilon}}{\Gamma\left(1-\epsilon\right)}\left(\frac{4\pi\mu_r^2}{s}\right)^{\epsilon}
%-----------------------
\nn\\
%-----------------------
&\times
C_A\Bigg\{\left(\frac{1}{2\epsilon^2}+\frac{17}{12}\frac{1}{\epsilon}-2\ln^2{\frac{\sqrt{r}}{1+\sqrt{r}}}-\frac{23}{6}\ln{\frac{\sqrt{r}}{1+\sqrt{r}}}-\frac{\pi^2}{4}+\frac{67}{36}\right)\delta(1+r-z)
%-----------------------
\nn\\
%-----------------------
& + \left[\frac{1}{1+r-z}\right]_{+}\left[2\ln{\frac{2-z+\sqrt{z^2-4r}}{2}}-\ln{\frac{z-\sqrt{z^2-4r}}{z+\sqrt{z^2-4r}}}\right.
%-----------------------
\label{dsigma:Div:dz}
\nn\\
%-----------------------
&\left. -\frac{2\sqrt{z^2-4r}}{1-r}+\frac{\sqrt{z^2-4r}\left(6+2r-6z+z^2\right)}{12(1-r)^3}\right]-\left[\frac{\ln{(1+r-z)}}{1+r-z}\right]_{+}\Bigg\},
%-----------------------
\\
%-----------------------
%-----------------------
%-----------------------
\frac{d\hat{\sigma}_{\text{Fin}}^{(8),gg}}{dz} =&
\hat{\sigma}_\text{LO}^{(8)}\frac{\alpha_s}{\pi}\frac{C_A}{12(1-r)^3(z-2r)^3(2-z)^2}\bigg\{\sqrt{z^2-4r}\,
\Big[ 16 r (3 - 9 r + 9 r^2 + 24 r^3 - 28 r^4 + 9 r^5)
%-----------------------
\nn\\
%-----------------------
& - 8 (3 - 3 r - 9 r^2 + 120 r^3 - 94 r^4 + 27 r^5) z +
 4 (6 - 15 r + 162 r^2 - 75 r^3 + 22 r^4) z^2
%-----------------------
\nn\\
%-----------------------
&- 2 (3 + 90 r + 23 r^2 + 4 r^3) z^3 +
 2 (12 + 25 r + 3 r^2) z^4 - (9 + 5 r) z^5 + z^6\Big]
%-----------------------
\nn\\
%-----------------------
&+ 12(1-r)^2\ln{\frac{z-2r-\sqrt{z^2-4r}}{z-2r+\sqrt{z^2-4r}}}\,\Big[ 4 r (1 - 2 r - 6 r^2 + 2 r^3 - 3 r^4) + 8 r^2 (6 + r + 3 r^2) z
%-----------------------
\nn\\
%-----------------------
&- 2 (1 + 12 r + 15 r^2 + 12 r^3) z^2 + 2 (3 + 9 r + 8 r^2) z^3 -
 2 (2 + 3 r) z^4 + z^5 \Big]\bigg\},
%-----------------------
\end{align}
%-----------------------
\eseq
%-----------------------
where $\hat{\sigma}_\text{LO}^{(8)}$ is given in Eq.~\eqref{hat:sigma:8:LO}.

From Eq.~\eqref{dsigma:Div:dz}, one immediately sees that the double and single IR poles indeed occur exactly
at the location $z=1+r$. The ``+''-function in Eq.~\eqref{dsigma:Div:dz} is understood in the distributive sense,
{\it i.e.},
%-----------------------
\begin{equation}
%%-----------------------
\label{eq:definition of the plus function}
%-----------------------
\int_{2\sqrt{r}}^{1+r}dz\, \left[f(z)\right]_{+}g(z)=\int_{2\sqrt{r}}^{1+r}dz\,f(z)\left[g(z)-g(1+r)\right],
%-----------------------
\end{equation}
%-----------------------
where $g(z)$ is an arbitrary test function that is regular at $z=1+r$.

Obtaining the analytic expressions in Eq.~\eqref{eq:differential cross section for gg8} requires more efforts than in the color-singlet channel, since double IR pole emerges in our case, whereas only single soft pole occurs in that case~\cite{Jia:2012qx}.
Some technical details about isolating IR singularities with DR method are expounded in Appendix~\ref{Three-body phase space integration}.

Further integrating Eq.~\eqref{eq:differential cross section for gg8} over the entire range of $z$,
we then obtain the integrated partonic cross section for $e^+ e^-\to c\bar{c}({}^1S_0^{(8)})+gg$:
%-----------------------
\begin{align}
%-----------------------
\label{cross section for gg}
%-----------------------
\hat{\sigma}_{\text{R}}^{(8),gg}=& \hat{\sigma}_\text{LO}^{(8)}\frac{\alpha_s}{\pi}
\frac{(1-r)^{-2\epsilon}r^{\epsilon}}{\Gamma\left(1-\epsilon\right)}
\left(\frac{4\pi\mu_r^2}{s}\right)^{\epsilon}C_A\left[\frac{1}{2\epsilon^2}+\frac{17}{12}\frac{1}{\epsilon}+\frac{2-3r}{16(1-r)}\ln^2{\frac{1-\sqrt{1-r}}{1+\sqrt{1-r}}}\right.
%-----------------------
\nn\\
%-----------------------
&\left.+\text{Li}_2\left(-\frac{1-r}{r}\right)+\frac{10-9r}{12(1-r)^{3/2}}\ln{\frac{1-\sqrt{1-r}}{1+\sqrt{1-r}}}+\frac{1-6\ln{r}}{6(1-r)}+\frac{235}{36}-\frac{5\pi^2}{12}\right].
%-----------------------
\end{align}
%-----------------------

We can carry out the real correction calculation for
$\gamma^{*}\to c\bar{c}(\fourIdx{1}{}{(8)}{0}{S})+q\bar{q}$ in a similar vein.
The squared amplitude in $d$ dimensions reads
%-----------------------
\begin{align}
%-----------------------
\label{eq:amplitude squared for 1S08 qq}
%-----------------------
\sum_{u,d,s}\sum_{\text{Pol,\,Col}} {\left|\mathcal{M}^{\rm R}\left[\gamma^{*}\to c\bar{c}(\fourIdx{1}{}{(8)}{0}{S})+q\bar{q}\right]\right|^2} = &
\frac{2^{10}\pi^3 e_c^2\alpha C_A C_F n_f \alpha_s^2\mu_r^{4\epsilon}(1-2\epsilon)}{(2-z)^2(1+r-z)s}
\notag\\
%-----------------------
&\times
\left[x_1^2+x_2^2-2(1+r-z)-\epsilon(z^2-4r)\right],
%-----------------------
\end{align}
%-----------------------
where $n_f=3$ represents the number of light flavors, where only $u$, $d$ and $s$ are retained.
The light quarks  are treated as massless.
After integrating Eq.~\eqref{eq:amplitude squared for 1S08 qq} over the energy fraction of
 the massless quark, $x_1$,
we obtain
%-----------------------
\begin{align}
%-----------------------
\label{eq:differential cross section for qqbar}
%-----------------------
\frac{d\hat{\sigma}_{\text{R}}^{(8),q\bar{q}}}{dz}=& \hat{\sigma}_\text{LO}^{(8)}\frac{\alpha_s}{\pi}
\frac{(1-r)^{-2\epsilon}r^{\epsilon}}{\Gamma\left(1-\epsilon\right)}\left(\frac{4\pi\mu_r^2}{s}\right)^{\epsilon}
%-----------------------
\frac{n_f}{6}\Bigg\{\left[-\frac{1}{\epsilon}+2\ln{\frac{\sqrt{r}}{1+\sqrt{r}}}-\frac{5}{3}\right]\delta(1+r-z)
%-----------------------
\nn\\
%-----------------------
&+\left[\frac{1}{1+r-z}\right]_{+}\frac{\left(z^2-4r\right)^{3/2}}{(1-r)(2-z)^2}\Bigg\}.
%-----------------------
\end{align}
%-----------------------
Unlike the case for $e^+ e^-\to c\bar{c}(\fourIdx{1}{}{(8)}{0}{S})+gg$, here only the single pole arises,
originating from the configuration where the light quark and antiquark become collinear.

The integrated expression for $ \hat{\sigma}_{\text{R}}^{(8),q\bar{q}} $ turns to be
%-----------------------
\beq
%-----------------------
\label{cross section for qqbar}
%-----------------------
\hat{\sigma}_{\text{R}}^{(8),q\bar{q}}=\hat{\sigma}_\text{LO}^{(8)}\frac{\alpha_s}{\pi}
\frac{(1-r)^{-2\epsilon}r^{\epsilon}}{\Gamma\left(1-\epsilon\right)}
\left(\frac{4\pi\mu_r^2}{s}\right)^{\epsilon}\frac{n_f}{6}\left(-\frac{1}{\epsilon}
-\frac{20-8r-9\ln{r}}{3(1-r)}-\frac{2}{\left(1-r\right)^{3/2}}\ln{\frac{1-\sqrt{1-r}}{1+\sqrt{1-r}}}\right).
%-----------------------
\eeq
%-----------------------

\subsection{Virtual correction}

In order to render finite predictions, one should further consider the virtual correction to $e^+e^-\to c\bar{c}({}^1S_0^{(8)})+g$,
which also contains IR singularities that serve to cancel
those IR singularities encountered in the real correction, as
encoded in Eqs.~\eqref{eq:differential cross section for gg8} and \eqref{eq:differential cross section for qqbar}.

One typical one-loop diagram is depicted in Fig.~\ref{fig:feynman diagrams}.
The partial fraction in the one-loop amplitudes is conducted with the aid of the package
\textsf{\$Apart}~\cite{Feng:2012iq}, and the integration-by-part reduction is facilitated by the package \textsf{FIRE}~\cite{Smirnov:2014hma}.
The resulting master integrals (MIs) are then calculated analytically,
against which are checked numerically by the package \textsf{LoopTools}~\cite{Hahn:1998yk}.
After the renormalization of the charm quark mass as well as the QCD coupling constant,
the UV divergences in the one-loop QCD amplitude will be eliminated.

Squaring the amplitudes and integrating over the two-body phase space, we obtain
%-----------------------
\begin{equation}
%-----------------------
\hat{\sigma}_\text{V}^{(8)}\equiv \int_{2\sqrt{r}}^{1+r}d z\frac{d\hat{\sigma}_{\text{V}}^{(8)}}{dz}=\frac{2\pi \alpha}{3s^2}\int{d\Phi_2}\,\sum_{\text{Pol,\,Col}}2\,{\rm Re}\bigg\{\mathcal{M}^{(0)\,*}
\left[\gamma^{*}\to c\bar{c}(\fourIdx{1}{}{(8)}{0}{S})+g\right]\mathcal{M}^{(1)}\left[\gamma^{*}\to c\bar{c}(\fourIdx{1}{}{(8)}{0}{S})+g\right]\bigg\},
%-----------------------
\end{equation}
%-----------------------
where $ \mathcal{M}^{(0)}$ denotes the tree-level amplitude for $ \gamma^{*}\to c\bar{c}(\fourIdx{1}{}{(8)}{0}{S})+g $,
and $\mathcal{M}^{(1)}$ represents the order-$\alpha_s$ one-loop QCD amplitude.
After substituting the analytical expressions for the MIs, and including the counterterm diagrams,
we are able to deduce the differential expression analytically for the virtual correction:
%-----------------------
\begin{align}\label{eq:differential cross section for virtual corrections}
%-----------------------
\frac{d\hat{\sigma}_{\text{V}}^{(8)}}{d z}=&\hat{\sigma}_\text{LO}^{(8)}\frac{\alpha_s}{\pi}\frac{(1-r)^{-2\epsilon}r^{\epsilon}}
{\Gamma\left(1-\epsilon\right)}\left(\frac{4\pi\mu_r^2}{s}\right)^{\epsilon}
\bigg\{-\frac{C_A}{2\epsilon^2}-\frac{2C_A+\beta_0}{4\epsilon}+\frac{\beta_0}{4}\ln{\frac{\mu_r^2}{m_c^2}}
%-----------------------
\nn\\
%-----------------------
&+\frac{C_A(2-r)-2C_F(2+r)}{8(1-r)}\ln^2{\frac{1-\sqrt{1-r}}{1+\sqrt{1-r}}}+\frac{3\left(C_A-2C_F\right)}{2\sqrt{1-r}}\ln{\frac{1-\sqrt{1-r}}{1+\sqrt{1-r}}}
%-----------------------
\nn\\
%-----------------------
&+\frac{C_A(1-r)+2C_F}{4(1-r)}\left(\ln^2{\frac{r}{2-r}}+2\text{Li}_2{\frac{r}{2-r}}\right)
%-----------------------
\nn\\
%-----------------------
& +\frac{-2C_A(2-r)+4C_F(3-2r)+(2-r)^2\beta_0}{2(2-r)^2}\ln{\frac{r}{2(1-r)}}+\frac{C_A\left(9+4\pi^2\right)}{6}
%-----------------------
\nn\\
& -\frac{C_F\left[\pi^2(2-r)+6(1-r)(9-5r)\right]}{6(2-r)(1-r)}\bigg\}\delta(1+r-z),
%-----------------------
\end{align}
%-----------------------
where $ \beta_0=\frac{11}{3}C_A-\frac{2}{3} n_f $ is the one-loop coefficient of the
QCD $\beta $-function, and $\mu_r$ refers to the renormalization scale.
Note here the $1/\epsilon^2$ and $1/\epsilon$ poles, which sit exactly at $z=1+r$, are entirely of infrared origin.

\subsection{Summing real and virtual corrections}

We proceed to infer the net NLO radiative correction to $e^+e^-\to \gamma^* \to c\bar{c}({}^1S_0^{(8)})+g$,
by adding up the real correction contributions, Eq.~\eqref{eq:differential cross section for gg8} from
the $c\bar{c}({}^1S_0^{(8)})+gg$ channel, Eq.~\eqref{eq:differential cross section for qqbar} from the
$c\bar{c}({}^1S_0^{(8)})+q\bar{q}$ channel, together with the virtual correction in Eq.~\eqref{eq:differential cross section for virtual corrections}:
%%-----------------------\frac{d\hat{\sigma}_\text{LO}^{(8)}}{d z}
%\beq
%%-----------------------
%{\color{red} \frac{d\hat{\sigma}^{\text{(1)}}_{8}}{dz}\equiv \frac{d\hat{\sigma}_{\text{NLO}}^{(8)}}{dz}-{d \hat{\sigma}_{\text{LO}}^{(8)}\over dz}}\equiv
%\frac{d\hat{\sigma}_{\text{R}}^{(8)}}{dz}+\frac{d\hat{\sigma}_{\text{V}}^{(8)}}{dz} = {d\hat{\sigma}_{\text{R}}^{(8),gg} \over dz}+ {d\hat{\sigma}_{\text{R}}^{(8),q\bar{q}}\over d z}+
%{d\hat{\sigma}_{\text{V}}^{(8)}\over dz}.
%%-----------------------
%\eeq
%%-----------------------
%-----------------------\frac{d\hat{\sigma}_\text{LO}^{(8)}}{d z}
\begin{equation}
%-----------------------
\frac{d\hat{\sigma}_{\text{NLO}}^{(8)}}{dz}\equiv \frac{d \hat{\sigma}_{\text{LO}}^{(8)}}{d z}+
\frac{d\hat{\sigma}_{\text{R}}^{(8)}}{dz}+\frac{d\hat{\sigma}_{\text{V}}^{(8)}}{dz} =\frac{d \hat{\sigma}_{\text{LO}}^{(8)}}{d z}+ {d\hat{\sigma}_{\text{R}}^{(8),gg} \over dz}+ {d\hat{\sigma}_{\text{R}}^{(8),q\bar{q}}\over d z}+
{d\hat{\sigma}_{\text{V}}^{(8)}\over dz}.
%-----------------------
\end{equation}
%-----------------------
As anticipated, all the double and single IR poles indeed cancel, and we end up with the differential NLO SDC for the color-octet
channel:
%-----------------------
\begin{align}
%-----------------------
\label{eq:dF8 NLO}
& \frac{dF_{8}^{\text{NLO}}}{dz}=\frac{dF_{8}^{\text{LO}}}{dz}+ F_{8}^{\text{LO}}\frac{\alpha_s}{\pi}\bigg\{\frac{\beta_0}{4}\ln{\frac{\mu_r^2}{m_c^2}}+\frac{C_A(2-r)-2C_F(2+r)}{8(1-r)}\ln^2{\frac{1-\sqrt{1-r}}{1+\sqrt{1-r}}}
%-----------------------
\nn\\
%-----------------------
&+\frac{3\left(C_A-2C_F\right)}{2\sqrt{1-r}}\ln{\frac{1-\sqrt{1-r}}{1+\sqrt{1-r}}}+\frac{C_A(1-r)+2C_F}{4(1-r)}\left(\ln^2{\frac{r}{2-r}}+2\text{Li}_2{\frac{r}{2-r}}\right)
%-----------------------
\nn\\
%-----------------------
&+\frac{-2C_A(2-r)+4C_F(3-2r)+(2-r)^2\beta_0}{2(2-r)^2}\ln{\frac{r}{2(1-r)}}-\frac{C_F\left[\pi^2(2-r)+6(1-r)(9-5r)\right]}{6(2-r)(1-r)}
\nn\\
%-----------------------
&-2C_A\ln^2{\frac{\sqrt{r}}{1+\sqrt{r}}}+\frac{-23C_A+2n_f}{6}\ln{\frac{\sqrt{r}}{1+\sqrt{r}}}+\frac{121C_A+15\pi^2C_A-10n_f}{36}\bigg\}\delta(1+r-z)
%-----------------------
\nn\\
%-----------------------
&+F_{8}^{\text{LO}}\frac{\alpha_s}{\pi}\Bigg\{\left[\frac{1}{1+r-z}\right]_{+}C_A\left[2\ln{\frac{2-z+\sqrt{z^2-4r}}{2}}-\ln{\frac{z-\sqrt{z^2-4r}}{z+\sqrt{z^2-4r}}}-\frac{2\sqrt{z^2-4r}}{1-r}\right.
%-----------------------
\nn
\\
%-----------------------
&\left.+\frac{\sqrt{z^2-4r}\left(6+2r-6z+z^2\right)}{12(1-r)^3}\right]-C_A\left[\frac{\ln{(1+r-z)}}{1+r-z}\right]_{+}+\frac{n_f}{6}\left[\frac{1}{1+r-z}\right]_{+}\frac{\left(z^2-4r\right)^{3/2}}{(1-r)(2-z)^2}\Bigg\}
%-----------------------
\nn
\\
%-----------------------
&+F_{8}^{\text{LO}}\frac{\alpha_s}{\pi}\frac{C_A}{12(1-r)^3(z-2r)^3(2-z)^2}\bigg\{\sqrt{z^2-4r}\,
\Big[ 16 r (3 - 9 r + 9 r^2 + 24 r^3 - 28 r^4 + 9 r^5)
%-----------------------
\nn\\
%-----------------------
& - 8 (3 - 3 r - 9 r^2 + 120 r^3 - 94 r^4 + 27 r^5) z +
4 (6 - 15 r + 162 r^2 - 75 r^3 + 22 r^4) z^2
%-----------------------
\nn\\
%-----------------------
&- 2 (3 + 90 r + 23 r^2 + 4 r^3) z^3 +
2 (12 + 25 r + 3 r^2) z^4 - (9 + 5 r) z^5 + z^6\Big]
%-----------------------
\nn\\
%-----------------------
&+ 12(1-r)^2\ln{\frac{z-2r-\sqrt{z^2-4r}}{z-2r+\sqrt{z^2-4r}}}\,\Big[ 4 r (1 - 2 r - 6 r^2 + 2 r^3 - 3 r^4) + 8 r^2 (6 + r + 3 r^2) z
%-----------------------
\nn\\
%-----------------------
&- 2 (1 + 12 r + 15 r^2 + 12 r^3) z^2 + 2 (3 + 9 r + 8 r^2) z^3 -
2 (2 + 3 r) z^4 + z^5 \Big]\bigg\},
%-----------------------
\end{align}
%-----------------------
where $F_8^{\rm LO}$ is given in Eq.~\eqref{eq:F8 LO}.
%{\color{red} The NLO coefficient
%$F_8^{\rm NLO}$ can be written as $F_8^{\rm NLO}\equiv F_8^{\rm LO}+F_{8}^{\text{(1)}}$. }

After integrating Eq.~\eqref{eq:dF8 NLO} over the entire range of $z$,
we then get the integrated NLO color-octet SDC:
%-----------------------
\begin{align}
%-----------------------
\label{eq:F8 NLO}
%-----------------------
F_{8}^{\text{NLO}}=&F_{8}^{\text{LO}}+  F_{8}^{\text{LO}}\frac{\alpha_s}{\pi}\bigg\{\frac{\beta_0}{4}
\ln{\frac{\mu_r^2}{m_c^2}}
+\frac{C_A(6-5r)-4C_F(2+r)}{16(1-r)}\ln^2{\frac{1-\sqrt{1-r}}{1+\sqrt{1-r}}}
%-----------------------
\nn\\
%-----------------------
& +\frac{C_A(28-27r)-36C_F(1-r)-4n_f}{12(1-r)^{3/2}}\ln{\frac{1-\sqrt{1-r}}{1+\sqrt{1-r}}}+C_A\text{Li}_2\left(-\frac{1-r}{r}\right)
%+\frac{C_A}{2}\left[2\text{Li}_2{r}-\ln^2{r}+2\ln{r}\ln{(1-r)}\right]
%-----------------------
\nn\\
%-----------------------
%-----------------------
&+\frac{C_A(1-r)+2C_F}{4(1-r)}\left(\ln^2{\frac{r}{2-r}}+2\text{Li}_2{\frac{r}{2-r}}\right)+\frac{C_A\left(1-6\ln{r}\right)}{6(1-r)}-\frac{n_f\left(20-8r-9\ln{r}\right)}{18(1-r)}
%-----------------------
\nn\\
%-----------------------
%-----------------------
&+\frac{-2C_A(2-r)+4C_F(3-2r)+(2-r)^2\beta_0}{2(2-r)^2}\ln{\frac{r}{2(1-r)}}
%-----------------------
\nn\\
%-----------------------
& -\frac{C_F\left[\pi^2(2-r)+6(1-r)(9-5r)\right]}{6(2-r)(1-r)}+\frac{C_A(289+9\pi^2)}{36}\bigg\}.
%-----------------------
\end{align}
%-----------------------

We note that the NLO radiative correction to $e^+e^-\to \gamma^*\to c\bar{c}({}^1S_0^{(8)})+g$ has already been
computed by Zhang {et al.}~\cite{Zhang:2009ym} about a decade ago.
Those authors employed a purely numerical recipe, and only presented the integrated partonic
cross section. In contrast, we have presented the analytical expressions for both differential and integrated NLO color-octet SDCs (see Eqs.~\eqref{eq:dF8 NLO} and \eqref{eq:F8 NLO}).
When taking the same input parameters, our numerical prediction from Eq.~\eqref{eq:F8 NLO} is consistent with theirs.

In the $ \sqrt{s}\gg m_c$ limit, the correction for the color-octet SDC $ \delta F_{8}\equiv F_{8}^{\text{NLO}}-F_{8}^{\text{LO}} $ reaches the following asymptotic form:
%-----------------------
\begin{align}
%-----------------------
\label{eq:Asymptotic behavior of F8 at NLO}
%-----------------------
\left.\delta F_{8}\right|_{\text{Asym.}}=& F_8^{\text{LO}}\frac{\alpha_s}{\pi}\left[\frac{\beta_0}{4}
\ln{\frac{\mu_r^2}{m_c^2}}+\frac{C_A}{8}\ln^2{r}-(2C_A-C_F)\ln{2}\ln{r} +\frac{16C_A-9C_F-n_f}{6}\ln{r} \right.
%-----------------------
\nn\\
%-----------------------
&+\left. \frac{7C_A-6C_F}{4}\ln^2{2}-\frac{12C_A-9C_F-2n_f}{2}\ln{2} +\frac{295C_A-162C_F-40n_f+3\left(C_A-2C_F\right)\pi^2}{36}\right].
%-----------------------
\end{align}
%-----------------------
In contrast to the asymptotic form of $F_1^{\text{LO}}(\mu_\Lambda)$
in Eq.~\eqref{eq:Asymptotic behavior of F1 at LO}, which is dominated by $\alpha_s^2 m_c \ln r/s^2$,
here it is the double logarithm $\ln^2{r}$ that accompanies the $\alpha_s^2 m_c/s^2$ factor.
Since $\ln^2 r \gg |\ln r|$ in asymptotically high energy, one may conclude that
the color-octet channel dominates the inclusive $h_c$ production rate over the color-singlet one, at sufficiently high energy.
The occurrence of $\ln^2 r$ at NLO strongly suggests that, in order to improve the reliability of the
fixed-order predictions,
it seems desirable to resum these types of double logarithms to all orders in $\alpha_s$ in the color-octet channel.
We believe that the appropriate formalism to achieve this goal is to combine double-parton fragmentation
approach~\cite{Fleming:2012wy,Kang:2014tta,Ma:2014svb} and NRQCD factorization, where the large logarithms can be resummed by
invoking the corresponding evolution equation.
Practically speaking, at $B$ factory energy, $\sqrt{s}\approx 10.6$ GeV,
$\ln r$ is not particularly large, so resummation does not sound absolutely necessary.
Nevertheless, in the next-generation $e^+e^-$ colliders, as exemplified by \textsf{CEPC} and \textsf{ILC},
with $\sqrt{s}\approx 250$ GeV, the logarithms become so huge that one is enforced to carry out this kind of resummation.

\section{Endpoint resummation for color-octet channel}\label{sec:resummation}

When we reach the endpoint region in which $z \to 1+r$ and the $h_c$ carries its maximally allowed energy, fixed-order calculations
are plagued with large endpoint logarithms of the form $ \sum_{j<i} \alpha_s^i \left[ \ln^{2i - j} (1+r - z)/(1+r-z)\right]_+ $.
This is clearly visible from those ``+'' distributions in our NLO color-octet prediction to
the $h_c$ energy spectrum in Eq.~\eqref{eq:dF8 NLO}.
To provide reliable predictions, these threshold logarithms have to be resummed to all orders. In this section we resum those logarithms to the NLL accuracy within the SCET framework~\cite{Bauer:2000ew,Bauer:2000yr,Bauer:2001ct,Bauer:2001yt,Bauer:2002nz,Beneke:2002ph}.

Following Ref.~\cite{Fleming:2003gt}, the factorization theorem for the color-octet $h_c$ production is found to take the form
%-----------------------
\beq
%-----------------------
\label{eq:factorization}
%-----------------------
\frac{d\sigma}{dz'}
= %p[r,z'] \,
\hat{\sigma}^{(8)}_\text{LO} \, H[\mu_H\,, \mu] \, \int_{z'}^1 d x S[x,\mu_S\,, \mu_r]
\times
J\left[ s(1+r)(x-z'), \mu_J\,, \mu_r \right] \,,
%-----------------------
\eeq
%-----------------------
where we have introduced
%-----------------------
\beq
%-----------------------
z' = \frac{E_{h_c}}{E_{h_c}^{\rm max}} = \frac{z}{1+r}.
%\quad\quad  p[r,z'] =  \frac{ \left[ (1+r)^2 z'^2 - 4 r  \right]^{\frac{1}{2}} }{1-r} \,.
%-----------------------
\eeq
%-----------------------
Here $H$ is the hard function normalized to $1$. The hard function which encodes the virtual corrections can be calculated perturbatively. Its one-loop results and
anomalous dimension $\gamma_H$ can be
extracted from Eq.~\eqref{eq:differential cross section for virtual corrections}. $S$ and $J$ stand for the shape and jet functions, respectively.

The shape function $S^{(8,^1S_0)}(\ell^+)$ is defined in terms of ultrasoft fields that carry ${\cal O}(\Lambda_{\text{QCD}})$ momentum:
%-----------------------
\beq
%-----------------------
S(\ell^+) = \frac{\langle0|\chi^\dag T^a\psi \, a^\dag_{h_c} a_{h_c} \delta(\ell^+ -i n\cdot D)\psi^\dag \, T^a\chi \, |0\rangle}{4m_c \braket{\mathcal{O}_8^{h_c}\left(\fourIdx{1}{}{}{0}{S}\right)}}.
\eeq
%-----------------------
Its normalizations is written as $\int d\ell^+ S(\ell^+)=1$.
The ultrasoft covariant derivative can be expressed as $D^\mu=\partial^\mu-ig_sA^\mu_{us}$ and the lightlike vectors are defined as $n^\mu=(1,0,0,-1)$ and $\bar{n}^\mu=(1,0,0,1)$.
$\chi$ and $\psi$ are the Pauli spinors as previously introduced in Eq.~\eqref{eq:NRQCD production operators}, and $a^\dag_{h_c} a_{h_c}$ is the projector to project onto the final $h_c$ state.

The jet function describes the collinear radiations recoil against the $h_c$ in the threshold region. The jet function is independent of the state of the charm quark-antiquark pair and is defined as
%-----------------------
\beq
%-----------------------
J(\bar{n}\cdot p n\cdot k+p^2_\perp) = -\frac{s(1+r)}{4\pi}{\rm Im} \left[i\int d^4y e^{ik\cdot y}
 \langle0|T\{\mathrm{Tr}[T^a B_\perp^{(0)\beta}(y)]
 \mathrm{Tr}[T^a B_{\perp \beta}^{(0)}(0)]\}|0\rangle\right],
%-----------------------
\eeq
%-----------------------
where the subscript $\perp$ denotes the perpendicular direction,  the superscript $(0)$ denotes the bare field, and $B_\perp^{\mu}$
is the collinear gauge invariant effective field which can be written as
%-----------------------
\beq
%-----------------------
B_\perp^{\mu} =\frac{1}{g_s}W^\dag({\cal P}_\perp^{\mu}+g_s(A^\mu_{n,q})_\perp)W,
%-----------------------
\eeq
%-----------------------
with a collinear gluon field $A^\mu_{n,q}$ and a collinear Wilson line $W_n(x)=\sum_{\mathrm{perms}}\exp\left(-g_s\frac{1}{{\cal \bar{P}}}\bar{n}\cdot A_{n,q}(x)\right)$. Here ${\cal P}$ is the projection operator which picks out the large component of the momenta to its right~\cite{Bauer:2000yr}.

The one-loop anomalous dimension $\gamma_J$ for the jet function can be found in Ref.~\cite{Fleming:2003gt}, while the anomalous dimension for the soft function can
be inferred from the consistency condition $\gamma_S  + \gamma_H + \gamma_J = 0$.

To resum the large endpoint logarithms, all components $H$, $J$ and $S$ in the factorization theorem will be evolved from their natural scales $\mu_H$, $\mu_J$ and $\mu_S$ to a common scale $\mu_r$ to evaluate the cross section, following the RGE
\beq
\frac{d F_i}{d \ln \mu} = \gamma_i \, F_i  \,,
\eeq
where $i$ runs over the hard ($H$), collinear ($J$) and the soft ($S$) modes. The scales $\mu_H$, $\mu_J$ and $\mu_S$ set the initial condition for the RG running and are chosen to minimize the logarithms in the higher order corrections to $H$, $J$ and $S$, respectively, which is found to be of order
%-----------------------
\begin{align}
%-----------------------
&\mu_H \sim \frac{s}{M}(1-r) \,, \quad \quad \mu_S \sim M \frac{1+r}{1-r} (1-z')  \,,
%-----------------------
\nn\\
%-----------------------
&
\mu_J \sim \sqrt{ \mu_H \mu_S}\,,
%-----------------------
\end{align}
%-----------------------
where we have introduced the mass of the heavy quark pair $M \equiv 2m_c$.
%The evolution can be achieved by solving the renormalization-group-equations $\mu \, \frac{ d F_i }{ d \mu } = \gamma_{F_i} F_i$ where $F_i$ denotes each function $H$, $J$ and $S$.
After combining all pieces and assuming $\mu_J =  \sqrt{ \mu_H \mu_S} $, we arrived at a compact form for the NLL cross section, which reads
%-----------------------
\beq
%-----------------------
\frac{ d \sigma_{\rm pert.}^{\rm NLL} }{d z'} =
%p[r,z'] \,
\sigma_\text{LO}^{(8)}
e^h \left[
\frac{\mu_H M}{s(1-r) }
\right]^{2 C_A A_\gamma [\mu_H,\mu_J]} \,
\left[
\frac{\mu_S }{M} \frac{1-r}{1+r}
\right]^\omega
\frac{e^{\omega\gamma_E}}{\Gamma[1-\omega]} \,
(1-z')^{-\omega} \,,
%-----------------------
\eeq
%-----------------------
where $\gamma_E$ is the Euler constant. We define the auxiliary parameters
%-----------------------
\begin{align}\label{eq:auxiliary parameters}
%-----------------------
h =&
2 C_A {\bar S}(\mu_H,\mu_r)  - A_H(\mu_H,\mu_r)
+2 C_A {\bar S}(\mu_S,\mu_r)
 - A_S(\mu_S,\mu_r)
- 4C_A {\bar S}(\mu_J,\mu_r) - A_J(\mu_J,\mu_r)  \,,
%-----------------------
\nn \\
%-----------------------
\omega =&  2 C_A A_\gamma[\mu_S\,, \mu_J] < 0.
%-----------------------
\end{align}
%-----------------------
In Eq.~\eqref{eq:auxiliary parameters}, ${\bar S}$ and $A_i$ are found to be
%-----------------------
\begin{subequations}
%-----------------------
\begin{align}
%-----------------------
{\bar S} (\mu_i,\mu_f) =& \left[
\frac{4\pi}{\alpha_s(\mu_i) } \,
\left(
1- \frac{1}{\rho} - \ln \rho
\right) \,
+ \frac{\beta_1}{2\beta_0} \ln^2 \rho
+(1-\rho+\ln \rho) \left(
\frac{\gamma_1}{\gamma_0} - \frac{\beta_1}{\beta_0}
\right)\right] \frac{  \gamma_0 }{4\beta_0^2},
%-----------------------
\\
%-----------------------
A_\gamma (\mu_i, \mu_f)
=& \frac{\gamma_0 }{2\beta_0}  \left[
\ln \rho + \frac{\alpha_s(\mu_i) }{4\pi} \left(
\frac{\gamma_1}{\gamma_0} - \frac{\beta_1}{\beta_0}
\right)  (\rho-1)
\right],
%-----------------------
\end{align}
%-----------------------
\end{subequations}
%-----------------------
where
%-----------------------
\begin{subequations}
%-----------------------
\begin{align}
%-----------------------
\rho =&\frac{\alpha_s(\mu_f)}{\alpha_s(\mu_i)},
%-----------------------
\\
%-----------------------
\beta_0 =& \frac{11}{3}C_A - \frac{2}{3} n_f \,, \quad \quad
 %-----------------------
 \\
 %-----------------------
\beta_1 =&
\frac{34}{3} C_A^2 - \frac{20}{3}C_A T_F n_f - 4 C_F T_F n_f\,,
 \\
%-----------------------
\gamma_0 =& 4\,,
\\
\gamma_1 =& \left( \frac{67}{9} - \frac{\pi^2}{3} \right) C_A - \frac{20}{9} T_F n_f \,.
%-----------------------
\end{align}
%-----------------------
\end{subequations}
%-----------------------
%\bea
%&&\rho=\frac{\alpha_s(\mu_f)}{\alpha_s(\mu_i)},
%\\
%&& \beta_0 = \frac{11}{3}C_A - \frac{2}{3} n_f \,, \quad \quad
%\nn \\
%&& \beta_1 =
%\frac{34}{3} C_A^2 - \frac{20}{3}C_A T_F n_f - 4 C_F T_F n_f\,,
%\nn \\
%&&\gamma_0 = 4 \,, \quad \quad \gamma_1 = \left( \frac{67}{9} - \frac{\pi^2}{3} \right) C_A - \frac{20}{9} T_F n_f \,.
%\eea

Up to NLL accuracy, $A_H$, $A_J$ and $A_S$ are obtained by truncating out the $\alpha_s$ term and replacing $\gamma_{0}$ in $A_\gamma$ with $\gamma^{H}_{0}$, $\gamma^{J}_{0}$ or $\gamma^{S}_{0}$:
%-----------------------
\begin{align}
%-----------------------
&\gamma_0^H = - \frac{34}{3}C_A +\frac{4}{3} n_f \,, \quad \quad
\gamma_0^J = 2 \beta_0 \,, \quad \quad
%-----------------------
\nonumber\\
%-----------------------
& \gamma_0^S =  - \gamma_H^0 - \gamma_0^J = 4 C_A \,.
\end{align}
%-----------------------
Last we note that when  $ (1-z') \sim {\cal O} \left(  \Lambda_{\rm QCD} /M  \right) $, the
process-independent shape function becomes non-perturbative and therefore a non-perturbative model $ {\cal S}_{\rm non-pert.}$ is required for describing the non-perturbative soft radiations and the resummed cross section is modified as
%-----------------------
\beq
%-----------------------
\frac{ d \sigma^{\rm NLL} }{dz'}
 = \int_{z'}^1 \frac{  d x  }{x}  \,
\frac{ d \sigma_{\rm pert.}^{\rm NLL} } {d x} \, {\cal S}_{\rm non-pert.}\left( \frac{z'}{x} \right)  \,,
%-----------------------
\eeq
%-----------------------
where the non-perturbative shape function is adopted by a modified version
of a model used in the decay of $B$ mesons~\cite{Leibovich:2002ys}
%-----------------------
\beq
%-----------------------
\label{eq:shape:function}
%-----------------------
{\cal S}_{\rm non-pert.}(\ell^+)
 =\frac{1}{\bar{\Lambda}}\frac{A^{A B}}{\Gamma(A B)}(x-1)^{A B-1}e^{-A(x-1)}\,,
%-----------------------
\eeq
%-----------------------
with $x=\ell^+/\bar{\Lambda}$ and $ \bar{\Lambda}\sim \mathcal{O}(\Lambda_{\text{QCD}}) $. Due to lack of data, the parameters $A$ and $B$ have large uncertainties.  However, the moments of the shape function can be expressed by the NRQCD operators and can be ordered by the power counting rules.
The $N$-th moment of the shape function is $\mathcal{O}(\Lambda^N_{\text{QCD}}) $. According to the above model for the non-perturbative shape function, we have

\begin{eqnarray}
%-----------------------
%-----------------------
\int_{\bar{\Lambda}}^\infty d \ell^+ {\cal S}_{\rm non-pert.}(\ell^+)&=&1\,,\\
\int_{\bar{\Lambda}}^\infty d \ell^+ {\cal S}_{\rm non-pert.}(\ell^+)\ell^+&=&\bar{\Lambda}(B+1)\,,\\
\int_{\bar{\Lambda}}^\infty d \ell^+ {\cal S}_{\rm non-pert.}(\ell^+)(\ell^+)^2&=&(\bar{\Lambda})^2(\frac{B}{A}+(B+1)^2)\,.
%-----------------------
\end{eqnarray}
%$\bar{\Lambda}=M_{h_c}-2m_c$.
Thus  the parameters $A$ and $B$  can be ordered as $A\sim B\sim \mathcal{O}(1)$. Future measurements shall be helpful to fit these two parameters.

\section{Numerical results
\label{sec:numerical results}}

In this section, we present the numerical predictions for the differential and integrated cross sections
for the inclusive $h_c$ production at the \textsf{Belle II} experiment, $\sqrt{s}=10.58$ GeV.
We adopt the running QED coupling constant $\alpha(\sqrt{s})=1/130.9$~\cite{Bodwin:2007ga}.
We take the the charm quark mass $ m_c=1.5~\text{GeV} $, and the characteristic hadronic
scale $ \Lambda_\text{QCD}=332~\text{MeV}$~\cite{Olive:2016xmw}.
For the process under consideration, we take the QCD renormalization scale $\mu_r= \sqrt{s}/2=5.29~\text{GeV}$,  and
take the default value of the strong coupling constant $\alpha_s(\mu_r)=\alpha_s(\sqrt{s}/2)=0.19 $, which is determined by the two-loop RGE formula~\cite{vanRitbergen:1997va,Chetyrkin:2000yt}.

\begin{figure}[htbp]
	\centering
	\includegraphics[width=0.85\linewidth]{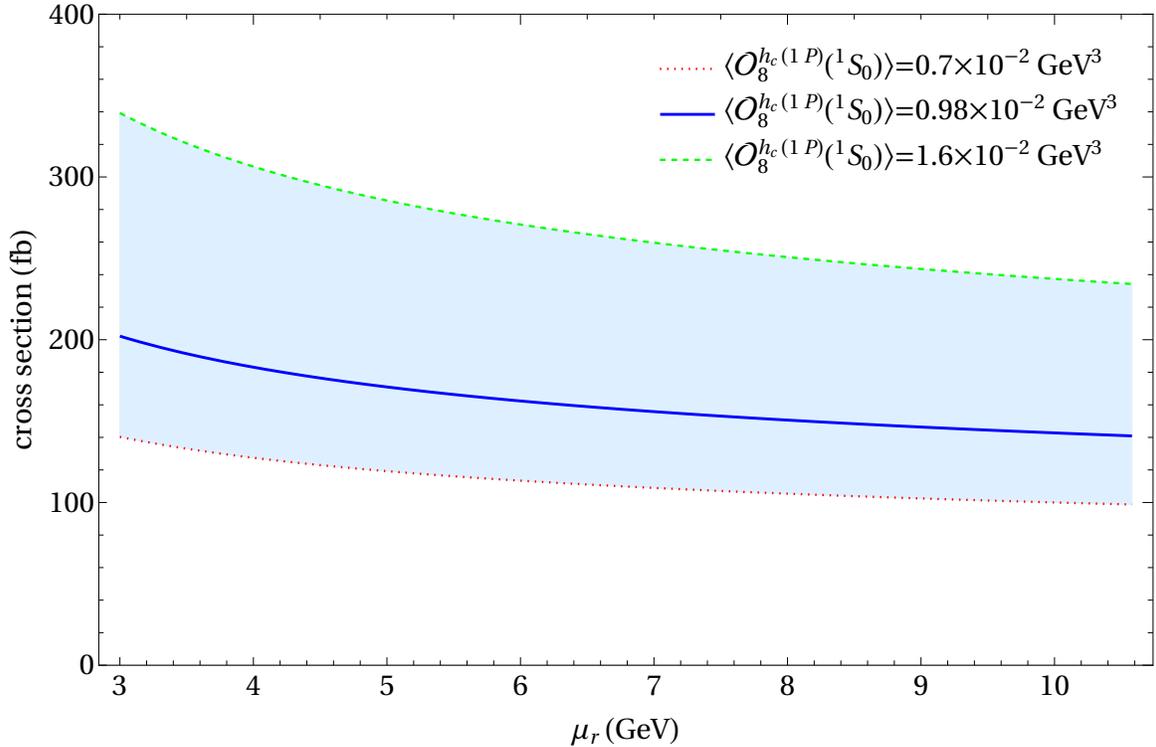}
	\caption{The dependence of total cross section $ \sigma_{\text{total}} $ on the renormalization
scale $ \mu_r $ and color-octet LDME $ \braket{\mathcal{O}_8^{h_c(1P)}\left(\fourIdx{1}{}{}{0}{S}\right)} $.
The scale $ \mu_r $ is varied from $ 2m_c $ to $ \sqrt{s} $ and the band represents the theoretical uncertainty due to the variation of the color-octet LDME.}
	\label{fig:totalcrosssectionchangingscaleando8}
\end{figure}

By definition in Eq.~\eqref{NRQCD formalism}, the total cross section $\sigma_{\text{total}}$ is obtained by summing the contributions from both color-singlet and octet channels, where the NLO QCD correction is included for the latter.
For the LDMEs in Eq.~\eqref{NRQCD formalism}, we take $\braket{\mathcal{O}_1^{h_c(1P)}\left(\fourIdx{1}{}{}{1}{P}\right)}=0.32~\text{GeV}^5$~\cite{Cho:1995vh,Bodwin:1994jh} in the color-singlet $h_c(1P)$ production.
In contrast, the color-octet LDME $ \braket{\mathcal{O}_8^{h_c(1P)}\left(\fourIdx{1}{}{}{0}{S}\right)}$ is poorly known,
which bears a large uncertainty.
In Table~\ref{tab:numerical results}, we present some benchmark choices for the color-octet LDME $ \braket{\mathcal{O}_8^{h_c(1P)}\left(\fourIdx{1}{}{}{0}{S}\right)} $ and the corresponding integrated cross sections from different channels.
%--------------------------------
In Fig.~\ref{fig:totalcrosssectionchangingscaleando8}, we also show the dependence of the integrated (total) cross section on the
renormalization scale $\mu_r$ and the color-octet LDME.
In other places of the paper, we will fix the value of this color-octet LDME as $ \braket{\mathcal{O}_8^{h_c(1P)}\left(\fourIdx{1}{}{}{0}{S}\right)}=0.98\times 10^{-2}~\text{GeV}^3 $~\cite{Cho:1995ce},
defined at the NRQCD factorization scale $\mu_\Lambda=m_c$.

%-----------------------
\begin{table}[htbp]
	\renewcommand\arraystretch{1.5}
	\begin{tabular}{c|c|c|c|c|c}
		\hline
		\hline
		Refs.& $ \braket{\mathcal{O}_8^{h_c(1P)}\left(\fourIdx{1}{}{}{0}{S}\right)}(\text{GeV}^3) $ & $ \sigma_{\text{LO}}^{(1)} $ & $ \sigma_{\text{LO}}^{(8)} $ & $ \sigma_{\text{NLO}}^{(8)} $ & $ \sigma_{\text{total}} $
		\\
		\hline
		\cite{Ma:2010vd,Gong:2012ug}& $ 0.7\times 10^{-2} $ &  & 69.41 & 127.09 & 117.36
		\\
		\cite{Cho:1995ce}& $ 0.98\times 10^{-2} $ & -9.73 & 97.17 & 177.92 & 168.20
		\\
		%\hline
		%\cline{1-2}\cline{4-5}
		\cite{Cho:1995vh}& $ 1.6\times 10^{-2} $ &  & 158.64 & 290.49 & 280.76
		\\
		%\hline
		%\cline{1-2}\cline{4-5}
		\hline
		\hline
	\end{tabular}
\caption{Numerical results for the integrated cross sections (fb) at $\sqrt{s}=10.58~\text{GeV}$, from various perturbative levels.
To assess the impact of the color-octet LDME on the cross sections,
we list different results by varying the value of $ \braket{\mathcal{O}_8^{h_c(1P)}\left(\fourIdx{1}{}{}{0}{S}\right)} $.}
\label{tab:numerical results}
\end{table}

%\label{totalcrosssectionchangingscaleando8}

From Table~\ref{tab:numerical results} and Fig.~\ref{fig:totalcrosssectionchangingscaleando8},
we find that the $h_c(1P)$ production cross section at $ \sqrt{s}=10.58~\text{GeV} $ is rather
sensitive to the color-octet LDME.
Therefore, the future measurements of the inclusive $h_c(1P)$ production at \textsf{Belle II} may provide a good place
to unearth the value of this color-octet LDME.

\begin{figure}[htbp]
	\centering
	\includegraphics[width=0.85\linewidth]{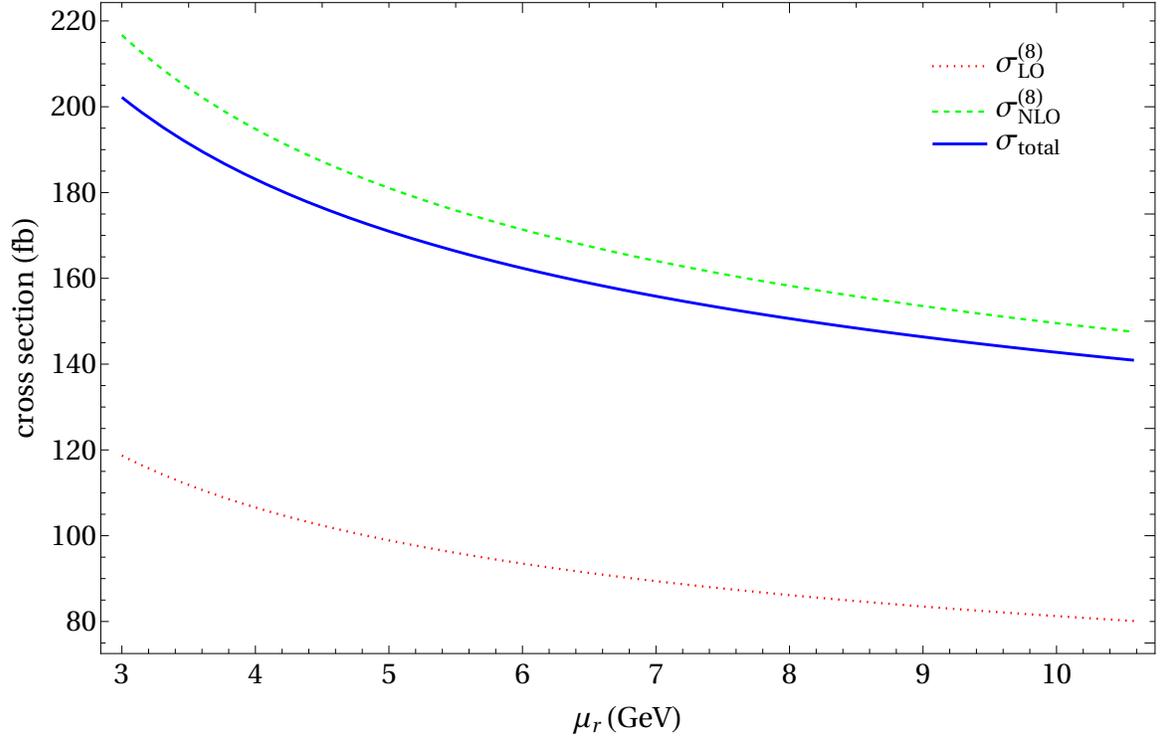}
	\caption{The dependence of the inclusive $ h_c(1P) $ production cross sections on the renormalization scale $\mu_r$,
 at various levels of perturbative orders. $\mu_r$ is ranging from $ 2m_c $ to $\sqrt{s}$.
 We have fixed the color-octet LDME $ \braket{\mathcal{O}_8^{h_c(1P)}\left(\fourIdx{1}{}{}{0}{S}\right)}=0.98\times 10^{-2}~\text{GeV}^3 $.}
	\label{fig:Total_Cross_Section_Scale}
\end{figure}

In Fig.~\ref{fig:Total_Cross_Section_Scale}, we also show the scale dependence of the integrated cross sections from each production channel, at various perturbative levels.
The scale $\mu_r$ is varied from $ 2m_c $ to $\sqrt{s}$.
From Table~\ref{tab:numerical results} and Fig.~\ref{fig:Total_Cross_Section_Scale}, one sees that the NLO QCD correction to the color-octet channel is important, with a $K$-factor of about $1.8$ and the color-octet contribution dominates the total production rate.
It is noteworthy that the color-singlet contribution in the $\overline{\rm MS}$ scheme even becomes negative.

\begin{figure}[htbp]
\centering
\includegraphics[width=0.85\linewidth]{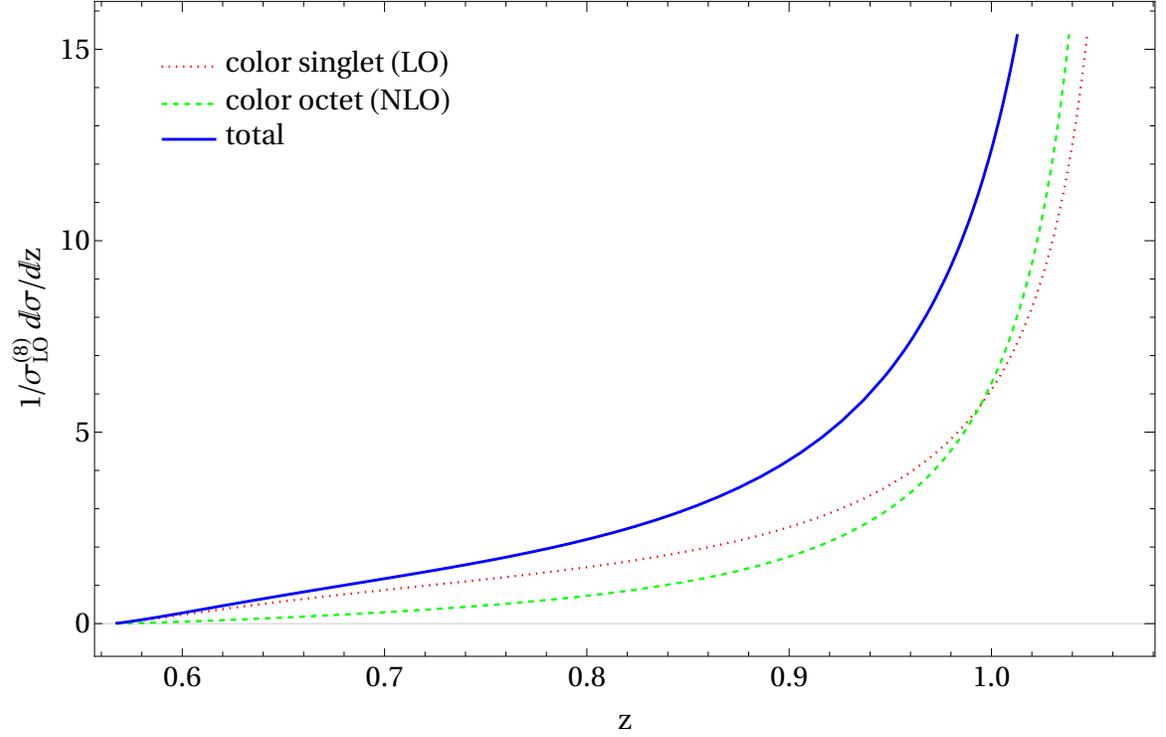}
\caption{The $h_c(1P)$ energy spectrum at $ \sqrt{s}=10.58~\text{GeV} $ from the fixed-order calculation.
In addition to their sum, we have also shown the contributions from the color-singlet channel at LO and the color-octet channel at NLO, all of which are normalized by the LO color-octet cross section $ \sigma_\text{LO}^{(8)} $.}
\label{fig:Fixed_Order_Distribution}
\end{figure}

To date, the \textsf{Belle I} experiment has accumulated an integrated luminosity about $711~\text{fb}^{-1}$ at $ \sqrt{s}=10.58~\text{GeV}$. Thus, from our calculation, around $(0.8-2)\times 10^5$ $h_c(1P)$ events should have already been produced.
Furthermore, we expect that roughly $(6-14)\times 10^6$ $ h_c(1P) $ events will be produced,
when the designed luminosity reaches $50~\text{ab}^{-1}$ at $ \sqrt{s}=10.58~\text{GeV}$ in the forthcoming \textsf{Belle II} experiment.

Such a huge data set of $h_c(1P)$ events may allow experimentalists to accurately
measure the $h_c(1P)$ differential energy spectrum.
The $h_c(1P)$ energy distribution from the fixed-order prediction is plotted in Fig.~\ref{fig:Fixed_Order_Distribution},
where the endpoint enhancement near $z\to 1+r$ can be readily visualized, indicating the breakdown of the fixed-order perturbative prediction near the maximal energy of $h_c(1P)$.

\begin{figure}[htbp]
	%----------------------------
	\centering
	%----------------------------
	\subfigure{\includegraphics[width=0.45\linewidth]{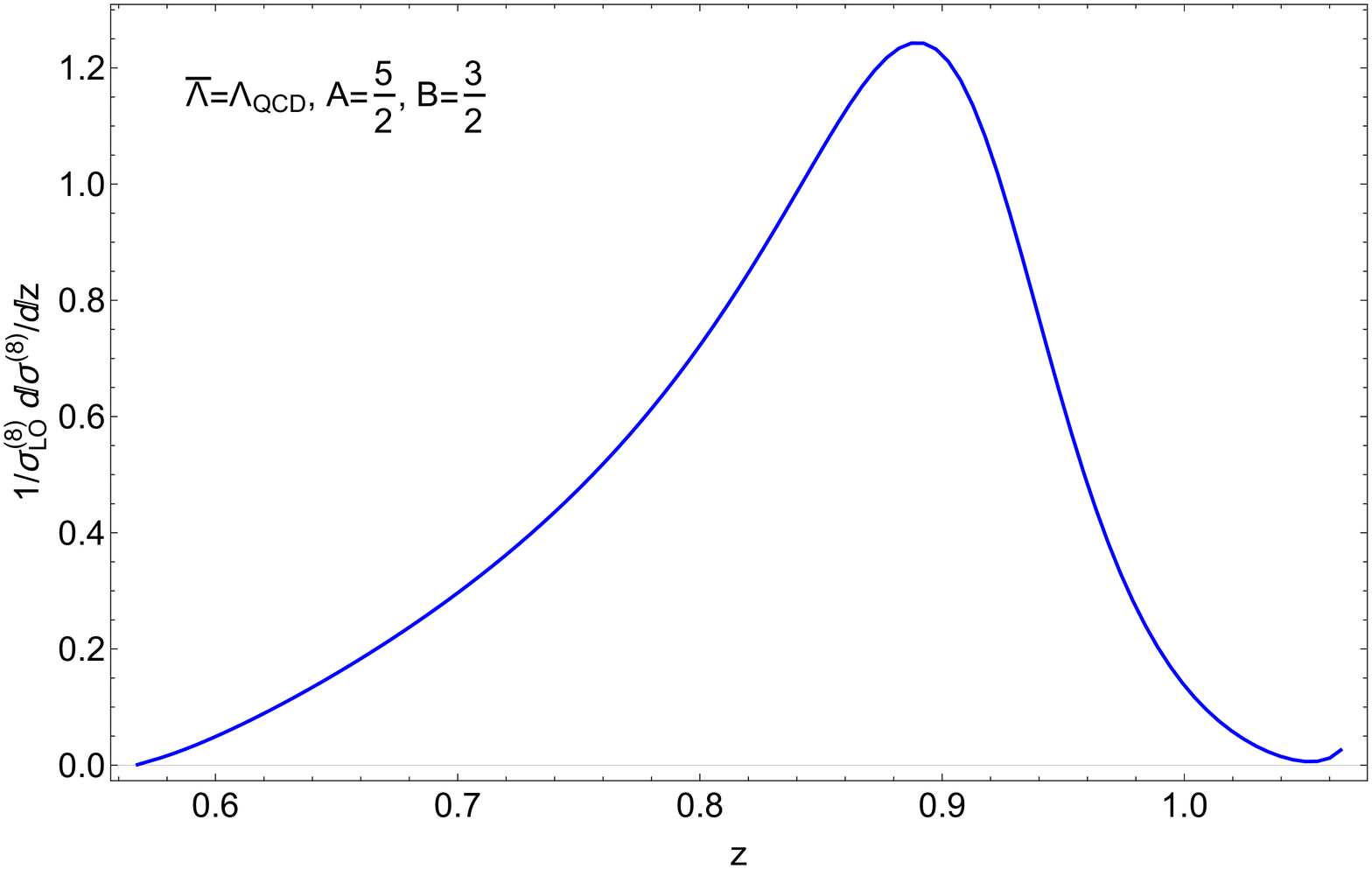}}
	\subfigure{\includegraphics[width=0.45\linewidth]{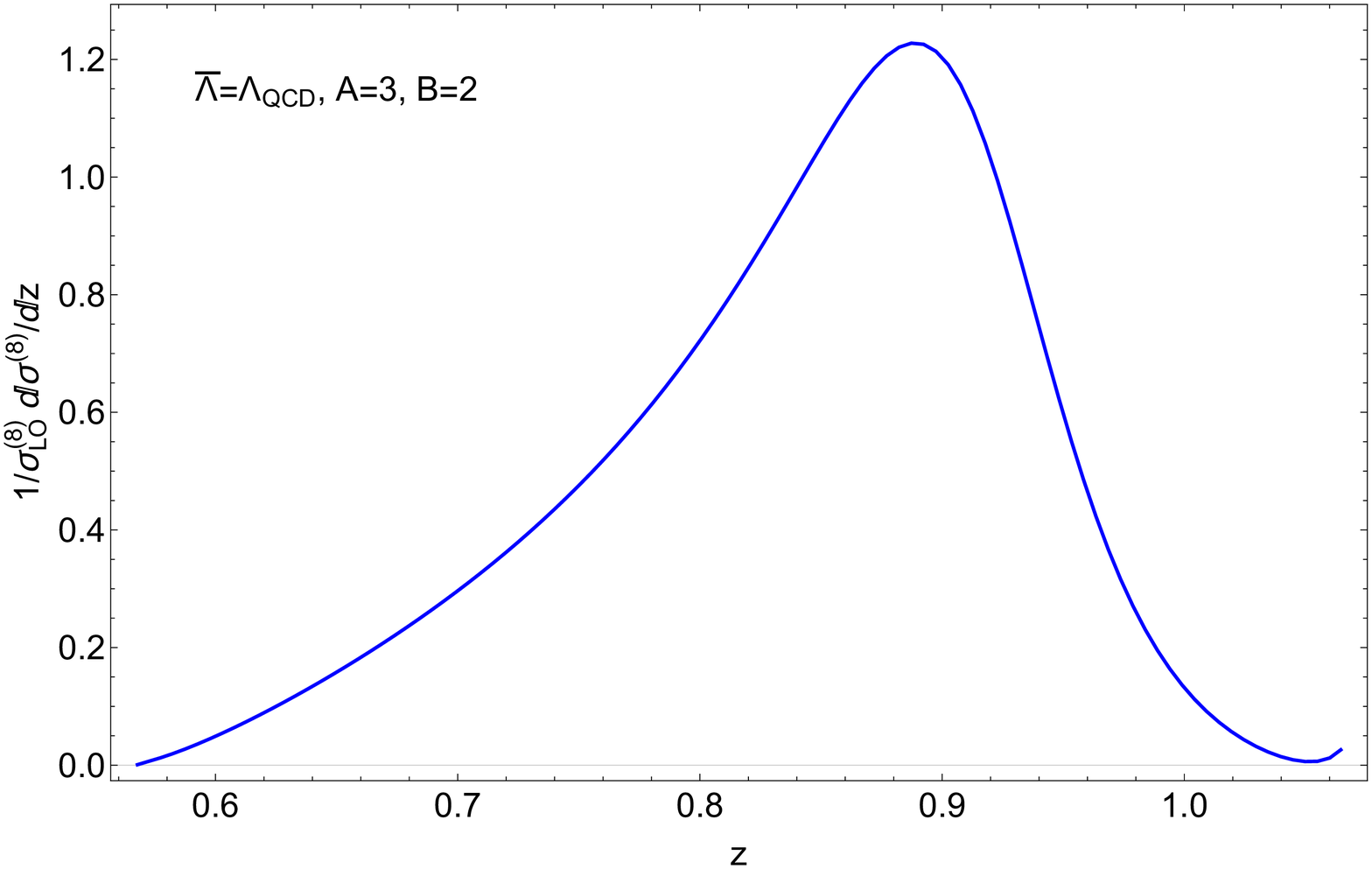}}
\subfigure{\includegraphics[width=0.45\linewidth]{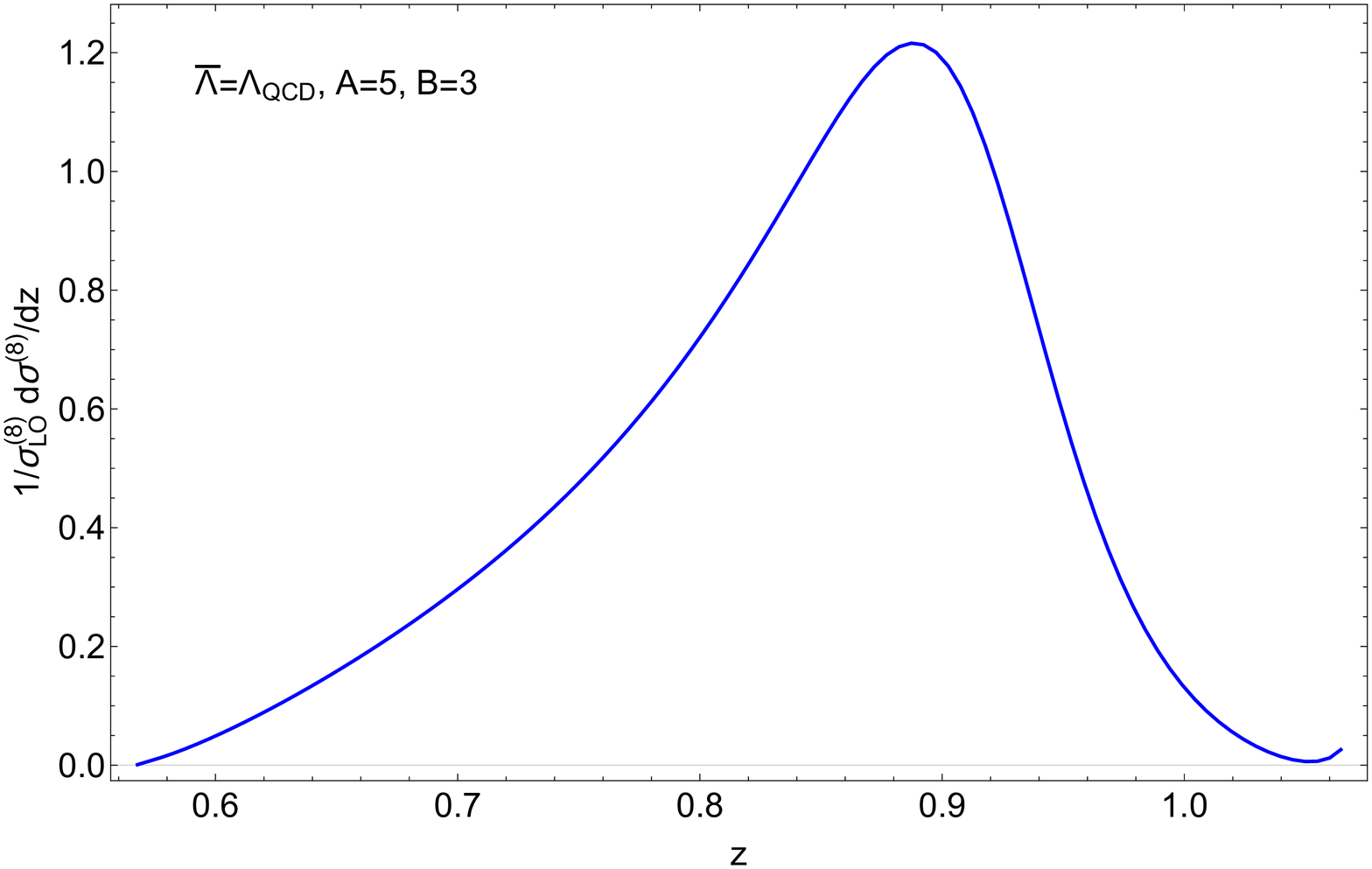}}
	\subfigure{\includegraphics[width=0.45\linewidth]{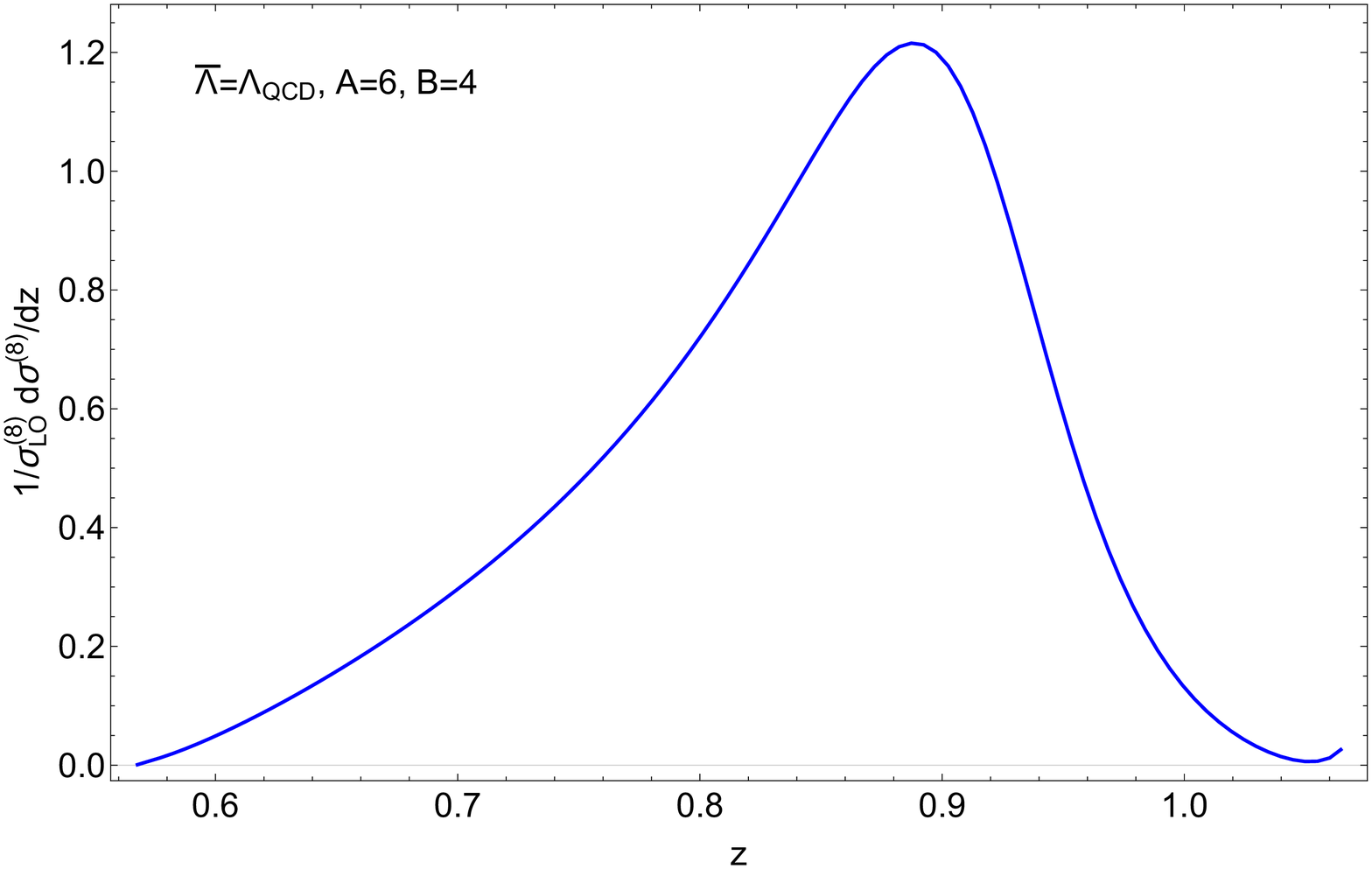}}
	%\caption{The solid curves are the NLO+NLL predictions for the $h_c$ energy spectrum, convoluted with shape function as given in
	%(\ref{shape:function}).}
	\caption{The NLO+NLL predictions for the $h_c(1P)$ energy spectrum solely from the color-octet channel.
	We have taken different numerical inputs for the parameters $ A $ and $ B $ first introduced in Eq.~\eqref{eq:shape:function}.}
	%----------------------------	
	\label{fig:Fixed_Order_Plus_NLL_With_Shape_Function}
\end{figure}

For the color-octet channel, the large endpoint logarithms have been resummed to the NLL accuracy within the SCET framework,
as expounded in Sec.~\ref{sec:resummation}, and the endpoint divergence problem can be resolved accordingly.
Away from the endpoint region, we merge the scales $ \mu_S=\mu_H =\mu_J = \mu_r $ to turn off the resummation effect.
While near the endpoint, we truncate the soft scale $\mu_S$ to around $1~\text{GeV}$, to avoid the Landau pole.
To account for the non-perturbative effects, we implement the  shape function model following
Ref.~\cite{Fleming:2003gt,Leibovich:2002ys}.  To merge all the scales to $\mu$ at small values of z, we adopted a ``profile function" which smoothly turns on resummation when z is small and turns off resummation by setting all the scales equal to $\mu$. The  profile function are chosen as $\frac{1\pm\tanh(15(z'-z'_{th.}))}{2}$~\cite{Liu:2013hba}. And the explicit form of $\mu_H(z)$ and $\mu_S(z)$ become as $\mu_H(z)=\frac{1-\tanh(15(z'-z'_{th.}))}{2}\mu_r+\frac{1+\tanh(15(z'-z'_{th.}))}{2}\frac{s}{2m_c}(1-r)$ and
$\mu_s(z)=\frac{1-\tanh(15(z'-z'_{th.}))}{2}\mu_r+\frac{1+\tanh(15(z'-z'_{th.}))}{2}$
where $z'=(1+r)z$ and $z'_{th.}$ is set to 0.85.  We further match the NLL resummation with the NLO results to obtain the
prediction for the full spectrum. The NLO $+$ NLL differential cross section is plotted in
Fig.~\ref{fig:Fixed_Order_Plus_NLL_With_Shape_Function},
where four sets of parameters for the shape function are adopted,  $(A=5/2,~B=3/2)$, $(A=3,~B=2)$, $(A=5,~B=3)$ and $(A=6,~B=4)$, respectively.
We can see that the unphysical enhancement near the kinematic endpoint is removed after taking the
resummation and shape function into account.

%----------------------------
%----------------------------

It is curious whether and how the $h_c(2P)$ meson, the first radially-excited spin-singlet $P$-wave charmonium,
could be observed at the Super $B$ factory.
To reconstruct the potential $ h_c(2P) $ events, one potentially useful decay chain is $h_c(2P)\to \eta_c(2S)\gamma$,
followed by $\eta_c(2S)\to h_c(1P)\gamma$, $h_c(1P)\to \eta_c\gamma$, and $\eta_c\to K^+ K^- \pi^0$. The decay chain, $h_c(2P)\to\eta_c(2S) \gamma$, followed by
$\eta(2S)\to K\overline{K}\pi$ may be another good channel for hunting
the $h_c(2P)$.
These decay channels are relatively clean, which hopefully will be helpful for
hunting the elusive $ h_c(2P) $ state.

The theoretical formulae for $ h_c(1P) $ can be readily transplanted to predict the inclusive production rate
of the $ h_c(2P)$ meson. We adopt the color-singlet LDME $ \braket{\mathcal{O}_1^{h_c(2P)}\left(\fourIdx{1}{}{}{1}{P}\right)}=0.438~\text{GeV}^5$~\cite{Jia:2012qx,Eichten:1995ch}.
It is rather difficult to accurately pin down the value of the color-octet LDME for $ h_c(2P)$,
and we follow the very rough estimation based on the RGE in Ref.~\cite{Bodwin:1994jh,Jia:2012qx},
and take $\braket{\mathcal{O}_8^{h_c(2P)}\left(\fourIdx{1}{}{}{0}{S}\right)}\approx 0.013~\text{GeV}^3$.
With these input parameters, and ignoring the small difference in phase space integration,
we then estimate the total cross section of $h_c(2P)$ to be around $224~\text{fb}$ at $ \sqrt{s}=10.58~\text{GeV}$.
When the integrated luminosity reaches $711~\text{fb}^{-1}$ ($50~\text{ab}^{-1}$) at this specific energy,
around $1.6\times 10^5$ ($1.1\times 10^7$) $ h_c(2P) $ events are expected to be produced.
The energy spectrum of the $h_c(2P)$ state assumes a similar shape as plotted in
Fig.~\ref{fig:Fixed_Order_Plus_NLL_With_Shape_Function}.

\section{Summary}
\label{sec:summary}

In this paper, we evaluate the NLO perturbative correction to the color-octet $h_c$ inclusive production
in $e^+e^-$ annihilation at the Super $B$ factory, within the NRQCD factorization framework.
We are able to deduce the analytic NLO color-octet SDC in a closed form.
The NLO correction from the color-octet channel is found to be positive and important.
Around $10^7$ $h_c(1P)$ and $h_c(2P)$ events are expected with the projected
$50~\text{ab}^{-1}$ luminosity at $ \sqrt{s}=10.58~\text{GeV}$
in the forthcoming \textsf{Belle II} experiment. It will be interesting to observe these $P$-wave spin-singlet
states in the inclusive production process.

Nevertheless, the $h_c$ energy spectrum predicted from the NLO calculation is plagued with the endpoint singularity,
which implies the failure of the fixed-order calculation near the maximal energy of $h_c$.
With the aid of the SCET formalism, these large endpoint logarithms are resummed to the NLL accuracy.
Consequently, in conjunction with the non-perturbative shape function, we obtain the well-behaved predictions for the
$h_c$ energy spectrum in the entire kinematic range, which are awaiting the close examination by
the forthcoming \textsf{Belle II} experiment.

%-----------------------
\section*{Acknowledgments}
%-----------------------
We are grateful to Cheng-Ping Shen for several useful discussions on experimental aspects.
%-----------------------
X. L. would like to thank the Kavli Institute of Theoretical Physics in Santa Barbara for the hospitality during the completion of this manuscript.
%-----------------------
This work was supported in part by the National Natural Science Foundation
of China under Grant No. 11375168, 11475188, 11621131001 (CRC110 by DGF and NSFC), and 11705092, by the Open Project Program of State Key Laboratory of Theoretical Physics under Grant No.~Y4KF081CJ1, by the IHEP Innovation Grant under contract number Y4545170Y2,
by the State Key Lab for Electronics and Particle Detectors.
%-----------------------

\begin{appendices}
	\renewcommand{\theequation}{\thesection.\arabic{equation}}
	\renewcommand{\thefigure}{\thesection.\arabic{figure}}
	\setcounter{equation}{0}
	\setcounter{figure}{0}
\section{Analytic integration over the three-body phase space}
\label{Three-body phase space integration}

In this Appendix, we explain how we derive the differential color-octet cross sections in Eq.~\eqref{eq:differential cross section for gg8} in DR.
Recall that the three-body phase space for $e^+e^-\to c\bar{c}({}^1S_0^{(8)})+gg(q\bar{q})$ in $ d=4-2\epsilon $ dimensions can be expressed as~\cite{Jia:2012qx}
%-----------------------
\begin{align}
%-----------------------
\label{eq:3 body phase space}
%-----------------------
\int{d\Phi_3}&=\frac{c_{\epsilon}(4\pi)^{\epsilon}}{\Gamma(1-\epsilon)}\left(\frac{s}{2}\right)^{1-2\epsilon}
\frac{1}{(4\pi)^3}\int_{2\sqrt{r}}^{1+r}dz\,\int_{a-b}^{a+b}dx_1\, x_1^{-2\epsilon}(z^2-4r)^{-\epsilon}\left(1-\cos^2{\theta}\right)^{-\epsilon}
%-----------------------
\nn\\
%-----------------------
&=\frac{c_{\epsilon}(4\pi)^{\epsilon}}{\Gamma(1-\epsilon)}\left(\frac{s}{2}\right)^{1-2\epsilon}\frac{1}{(4\pi)^3}\int_{2\sqrt{r}}^{1+r}dz\,\int_{a-b}^{a+b}dx_1\, 2^{-2\epsilon}(1+r-z)^{-\epsilon}\left(x_1-a+b\right)^{-\epsilon}\left(a+b-x_1\right)^{-\epsilon}
%-----------------------
\nn\\
%-----------------------
&=\frac{c_{\epsilon}(4\pi)^{\epsilon}}{\Gamma(1-\epsilon)}\frac{s^{1-2\epsilon}}{2}\frac{1}{(4\pi)^3}\int_{2\sqrt{r}}^{1+r}dz\, (1+r-z)^{-\epsilon}\int_{-b}^{b}d{\eta}\, \left(b+\eta\right)^{-\epsilon}\left(b-\eta\right)^{-\epsilon},
%-----------------------
\end{align}
%-----------------------
where $c_\epsilon$ is introduced in Eq.~\eqref{Def:cepsilon:and:r}, and $\theta $ is the polar angular between $ {\bf k}_1$ and $ \bf{P} $:
%-----------------------
\beq
\cos{\theta}=\frac{2(1+r-z)-x_1(2-z)}{x_1\sqrt{z^2-4r}}.
\eeq
%-----------------------
In Eq.~\eqref{eq:3 body phase space}, we have introduced three auxiliary variables $ a $, $ b $ and $ \eta $:
%-----------------------
\begin{equation}
a=\frac{2-z}{2}, \quad b=\frac{\sqrt{z^2-4r}}{2} \quad \text{and} \quad \eta=x_1-a,
%-----------------------
\end{equation}
%-----------------------
which satisfies $ a^2-b^2=1+r-z $.

First, let us concentrate on the soft term $\mathcal{I}_{\text{S}}\left(x_i,z\right) $  in Eq.~\eqref{eq:soft term}.
Upon integrating over energy fraction of gluon 1, it will result in a single IR pole.
For the sake of clarity, we discard the irrelevant perfectors in Eq.~\eqref{eq:3 body phase space},
and consider the following integral:
%-----------------------
\begin{align}
%-----------------------
\label{eq:AS}
%-----------------------
\mathcal{A}_{\text{S}} \equiv & \int_{a-b}^{a+b}dx_1\,\frac{(1+r-z)^{-\epsilon}\left(x_1-a+b\right)^{-\epsilon}\left(a+b-x_1\right)^{-\epsilon}}{\left(1+r-z-x_1\right)^2},
\notag\\
=&\frac{1}{(1+r-z)^{1+2\epsilon}}\int_{a-b}^{a+b}dt\, \frac{(1-\frac{a-b}{t})^{-\epsilon}(\frac{a+b}{t}-1)^{-\epsilon}}{(1-t)^2}.
%-----------------------
\end{align}
%-----------------------
In the second line, we change the integration variable from $x_1$ to $t$~\cite{Jia:2012qx},
%-----------------------
\begin{equation}
%-----------------------
%t=\frac{1+r-z}{x_1}=a+b \cos \theta,
t=\frac{1+r-z}{x_1},
%-----------------------
\end{equation}
%-----------------------
which lies in the range
%-----------------------
\begin{equation}
0<a-b<t<a+b<1-r.
\end{equation}
%-----------------------	

To explicitly identify the IR pole in $ \mathcal{A}_\text{S} $, we can rewrite~\cite{Jia:2012qx}
%-----------------------
\begin{equation}
%-----------------------
\label{eq:identity about distributions}
%-----------------------
\frac{1}{(1+r-z)^{1+2\epsilon}}=-\frac{\delta(1+r-z)}{2\epsilon\left(1-\sqrt{r}\right)^{4\epsilon}}+\left[\frac{1}{1+r-z}\right]_{+}
%-----------------------
-2\epsilon\left[\frac{\ln{(1+r-z)}}{1+r-z}\right]_{+}+\mathcal{O}(\epsilon^2),
%-----------------------
\end{equation}
%-----------------------
where the ``+''-function is defined in Eq.~\eqref{eq:definition of the plus function}.

Now the integration over $t$ in Eq.~\eqref{eq:AS} is convergent, therefore one can expand the integrand in powers of $\epsilon$. Through the order-$\epsilon^0$, $ \mathcal{A}_S $ bears the following form:
%-----------------------
\bqa
%-----------------------
\label{eq:AS 2}
%-----------------------
\mathcal{A}_\text{S}=\bigg\{-\frac{1-r}{2r\epsilon}+\frac{(1-r)
\left[4\ln{\left(1-\sqrt{r}\right)}+\ln{r}\right]}{2r}\bigg\}\delta(1+r-z)+
\left[\frac{1}{1+r-z}\right]_{+}\frac{\sqrt{z^2-4r}}{r}+\mathcal{O}(\epsilon).
%-----------------------
%-----------------------
\eqa
%%-----------------------

The soft-collinear term $ \mathcal{I}_{\text{SC}}\left(x_i,z\right) $ in
Eq.~\eqref{eq:soft and collinear term} would result in double IR pole upon phase space integration.
To facilitate the extraction of the IR poles, we first observe that $ \mathcal{I}_{\text{SC}}$
contains the following term:
%-----------------------
\beq
%-----------------------
\label{separation of ASC}
%-----------------------
\frac{1}{(1+r-z)\left(1+r-z-x_1\right)}=\frac{1}{x_1\left(1+r-z-x_1\right)}-\frac{1}{x_1\left(1+r-z\right)},
%-----------------------
\eeq
%-----------------------
which can be decomposed into two pieces through partial fraction.

The first term in Eq.~\eqref{separation of ASC} only leads to soft singularity.
Following the trick of changing variable in Eq.~\eqref{eq:AS}, we can readily work out the following
integration in DR:
%-----------------------	
\begin{align}
%-----------------------	
\label{eq:AS'}
%-----------------------	
\mathcal{A}'_\text{S}=&\int_{a-b}^{a+b}dx_1\,\frac{(1+r-z)^{-\epsilon}\left(x_1-a+b\right)^{-\epsilon}\left(a+b-x_1\right)^{-\epsilon}}{x_1\left(1+r-z-x_1\right)}
%-----------------------	
\nn\\
%-----------------------	
=&\frac{1}{(1+r-z)^{1+2\epsilon}}\int_{a-b}^{a+b}dt\,\frac{(1-\frac{a-b}{t})^{-\epsilon}(\frac{a+b}{t}-1)^{-\epsilon}}{t-1}
%-----------------------	
\nn\\
%-----------------------	
=&\bigg\{-\frac{\ln{r}}{2\epsilon}+\frac{\ln{r}}{4}\left[\ln{r}+8\ln{\left(1-\sqrt{r}\right)}\right]\bigg\}\delta{(1+r-z)}+\left[\frac{1}{1+r-z}\right]_{+}\ln{\frac{z-\sqrt{z^2-4r}}{z+\sqrt{z^2-4r}}}+\mathcal{O}(\epsilon).
%-----------------------	
\end{align}
%-----------------------

The second term in Eq.~\eqref{separation of ASC} would lead to double IR pole upon integration over $x_1$.
We then face the following integral:
%-----------------------	
\begin{align}
%-----------------------	
\label{JSC1}
%-----------------------	
\mathcal{A}_{\text{SC}}=&\int_{a-b}^{a+b}dx_1\,\frac{(1+r-z)^{-\epsilon}\left(x_1-a+b\right)^{-\epsilon}\left(a+b-x_1\right)^{-\epsilon}}{x_1\left(1+r-z\right)}
%-----------------------	
\nn\\
%-----------------------	
=&\frac{1}{(1+r-z)^{1+\epsilon}}\int_{a-b}^{a+b}dx_1\,\frac{\left(x_1-a+b\right)^{-\epsilon}\left(a+b-x_1\right)^{-\epsilon}}{x_1}
%-----------------------	
\nn\\
%-----------------------	
=&\frac{1}{(1+r-z)^{1+\epsilon}}\int_{-b}^{b}d\eta\,\frac{\left(\eta+b\right)^{-\epsilon}\left(b-\eta\right)^{-\epsilon}}{\eta+a}.
%-----------------------	
\end{align}
%-----------------------	
In the last step, we have switched the integration variable from $x_1$ to $\eta$,
as specified in the last line of Eq.~\eqref{eq:3 body phase space}.
%Note the prefactor outside the integral in \eqref{JSC1} can be rewritten in the form of distributions
%according to the identity \eqref{eq:identity about distributions}, which contributes a single
%IR pole located at $z=1+r$.

The integration over $\eta$ can be done in a straightforward way:
%-----------------------	
\begin{align}
%-----------------------	
\label{JSC2}
%-----------------------	
&\int_{-b}^{b} d\eta\,\frac{\left(\eta+b\right)^{-\epsilon}\left(b-\eta\right)^{-\epsilon}}{\eta+a}=
\frac{\sqrt{\pi}b^{1-2\epsilon}\Gamma(1-\epsilon){\fourIdx{}{2}{}{1}{F}\left(\frac{1}{2},1;\frac{3}{2}-\epsilon ;\frac{b^2}{a^2}\right)}}{a\Gamma\left(\frac{3}{2}-\epsilon\right)}
%-----------------------	
\nn\\
%-----------------------	
&=\frac{\sqrt{\pi}b^{1-2\epsilon}\Gamma(-\epsilon)\,{\fourIdx{}{2}{}{1}{F}\left(\frac{1}{2},1;1+\epsilon ;1-\frac{b^2}{a^2}\right)}}{a\Gamma\left(\frac{1}{2}-\epsilon\right)}
+\pi \csc{(\pi\epsilon)}\left(a^2-b^2\right)^{-\epsilon},
%-----------------------	
\end{align}
%-----------------------	
where ${}_2F_1$ represents the Gauss hypergeometric function. The singularity associated with the $\epsilon\to 0$ limit
can be readily traced in this format, which stems from $\Gamma(-\epsilon)$ and $\csc(\pi\epsilon)$.
With the aid of the package \textsf{HypExp}~\cite{Huber:2007dx},
we find the following expansion formula particularly useful:
%-----------------------	
\begin{equation}
%-----------------------	
\label{expand 2F1}
%-----------------------	
	{\fourIdx{}{2}{}{1}{F}\left(\frac{1}{2},1;1+\epsilon ;1-\frac{b^2}{a^2}\right)}=\frac{a}{b}\left(1+2\epsilon \ln{\frac{2b}{a+b}}\right)+\mathcal{O}{(\epsilon^2)}.
%-----------------------	
\end{equation}
%-----------------------	
%In Eq.~\eqref{expand 2F1}, the $ \mathcal{O}{(\epsilon^2)} $ term will finally vanish due to the $ \delta $ function in %Eq.~\eqref{eq:identity about distributions}, so it is legitimate to expand the hypergeometric function to $ \mathcal{O}{(\epsilon)} $.

Combining the distribution identity in Eq.~\eqref{eq:identity about distributions}, we can get
%-----------------------
\bseq
%-----------------------
\label{eq:ASC results}
%-----------------------
\begin{align}
%-----------------------
&\frac{1}{(1+r-z)^{1+\epsilon}}\frac{\sqrt{\pi}b^{1-2\epsilon}\Gamma(-\epsilon){\fourIdx{}{2}{}{1}{F}
\left(\frac{1}{2},1;1+\epsilon ;1-\frac{b^2}{a^2}\right)}}{a\Gamma\left(\frac{1}{2}-\epsilon\right)}
\nn\\
%-----------------------
=&\bigg\{\frac{1}{\epsilon^2}-\frac{2\ln{\left[(1-\sqrt{r})(1-r)\right]}}
{\epsilon}+2\ln^2{\left[(1-\sqrt{r})(1-r)\right]}-\frac{\pi^2}{6}\bigg\}\delta(1+r-z)
%-----------------------
\nn\\
%-----------------------
&+\left(-{1 \over \epsilon}+2\ln {2-z+\sqrt{z^2-4r} \over 2}\right)
\left[\frac{1}{1+r-z}\right]_{+}+\left[\frac{\ln(1+r-z)}{1+r-z}\right]_{+}+\mathcal{O}(\epsilon),
%-----------------------
\\
%-----------------------
&\frac{1}{(1+r-z)^{1+\epsilon}}\pi \csc(\pi \epsilon)\left(a^2-b^2\right)^{-\epsilon}
%-----------------------
\\
%-----------------------	
=&\bigg\{-\frac{1}{2\epsilon^2}+\frac{2\ln\left(1-\sqrt{r}\right)}
{\epsilon}-4\ln^2\left(1-\sqrt{r}\right)-\frac{\pi^2}{12}\bigg\}\delta(1+r-z)+\frac{1}{\epsilon}\left[\frac{1}{1+r-z}\right]_{+}-2\left[\frac{\ln(1+r-z)}{1+r-z}\right]_{+}+\mathcal{O}(\epsilon).
%-----------------------
\nn
%-----------------------
\end{align}
%-----------------------
\eseq
%-----------------------
Adding these two pieces together, we arrive at the final expression for $ \mathcal{A}_{\text{SC}} $:
%-----------------------
\begin{align}
%-----------------------	
\label{eq:ASC results final}
%-----------------------	
\mathcal{A}_{\text{SC}}=& \bigg\{\frac{1}{2\epsilon^2}-\frac{2\ln(1-r)}{\epsilon}-4\ln^2 \left(1-\sqrt{r}\right)	+2\ln^2{\left[(1-\sqrt{r})(1-r)\right]}-\frac{\pi^2}{4}\bigg\} \delta(1+r-z)
%-----------------------		
\notag\\
%-----------------------	
&+2\ln{\left(\frac{2-z+\sqrt{z^2-4r}}{2}\right)}\left[\frac{1}{1+r-z}\right]_{+}-
\left[\frac{\ln{(1+r-z)}}{1+r-z}\right]_{+}+\mathcal{O}(\epsilon).
%-----------------------	
\end{align}
%-----------------------
The occurrence of double IR pole is as anticipated, by examining
the pole structure of Eq.~\eqref{eq:soft and collinear term}.

\end{appendices}
%-----------------------

%-----------------------

\end{document}